\documentclass[letterpaper, 10 pt, conference]{ieeeconf}  % Comment this line out if you need a4paper

\IEEEoverridecommandlockouts                              % This command is only needed if 
\usepackage{mathrsfs}
\usepackage{bm}

\usepackage{graphicx}
\usepackage{wrapfig,lipsum}
\usepackage{pgfplots}
\usepackage{amsmath,amssymb,amsfonts}  
\usepackage{mathtools}
\usepackage{tikz, tikzscale}
\usepackage{caption}
\usepackage{subcaption}
\usepackage{enumerate}
\usepackage{comment}
\usepackage{booktabs}
\usepackage{dsfont}
\newtheorem{lem}{Lemma}
\newtheorem{proposition}{Proposition}
\newtheorem{corollary}{Corollary}
\newtheorem{example}{Example}
\newcommand{\qed}{\hfill $\blacksquare$} % For a solid black square

\usepackage{pbalance}

\newtheorem{defin}{Definition}
\newtheorem{ass}{Assumption}
\newtheorem{thm}{Theorem}

\newenvironment{lemma}{\begin{lem}}{\hfill  \end{lem}}
\newenvironment{definition}{\vskip 3pt\begin{defin}}{\vskip 3pt%% \hfill $\square$ 
\end{defin}}

\newcommand{\Gs}{\mathcal{G}_s}

\newcommand{\rednotes}[1]{\textcolor{red}{#1}} 
 
\newcommand{\magentanotes}[1]{\textcolor{magenta}{#1}} 

\title{\LARGE \bf
Topology Identification of Directed Signed Networks\\ via Repelling Laplacians
}

\author{Muhammad Ahsan Razaq \quad Pelin \c{S}ekercio\u{g}lu \quad Angela Fontan \quad Dimos V. Dimarogonas \quad Claudio Altafini % <-this % stops a space
\thanks{*This work was supported in part by the Wallenberg AI, Autonomous Systems and Software Program (WASP) funded by the Knut and Alice Wallenberg (KAW) Foundation, the Swedish Research Council (VR), Digital Futures, and by the ELLIIT framework program at Link\"oping University. }
\thanks{M. A. Razaq and C. Altafini are with the Division of Automatic Control, Department of Electrical Engineering, Linkoping University, SE-58183 Linkoping, Sweden. email: {\tt\small \{muhammad.ahsan.razaq, claudio.altafini\}@liu.se}. P. \c{S}ekercio\u{g}lu, A. Fontan, and D. V. Dimarogonas are with the Department of Decision and Control Systems, KTH Royal Institute of Technology, SE-100 44 Stockholm, Sweden. email:
        {\tt\small \{pelinse,angfon,dimos\}@kth.se} Corresponding author: P. Sekercioglu (email:
        {\tt\small pelinse@kth.se}).}%
}
\begin{document}

\maketitle
\thispagestyle{empty}
\pagestyle{empty}

%%%%%%%%%%%%%%%%%%%%%%%%%%%%%%%%%%%%%%%%%%%%%%%%%%%%%%%%%%%%%%%%%%%%%%%%%%%%%%%%
\begin{abstract}
We address the problem of simultaneously identifying the interaction topology and synchronizing a network interconnected over a directed graph that exhibits both cooperative and antagonistic interactions, modeled via a repelling Laplacian matrix. We propose an adaptive control protocol built on a node-centric regressor formulation, in which the network is driven to track a persistently exciting auxiliary network. We then propose a time-varying adaptive gain to remove the need for global knowledge of the network. We establish uniform semiglobal practical asymptotic stability of the estimation errors when the excitation depends only on measurable errors. Numerical simulations validate the theoretical results.
%We propose an adaptive control protocol for identifying the interaction topology of dynamical networks interconnected over directed graphs that exhibit both cooperative and antagonistic interactions, where the signed network is modeled via a repelling Laplacian. The proposed approach formulates the topology identification problem from an edge-based approach and leverages adaptive control through the design of a persistently exciting auxiliary network. This enables the simultaneous identification and synchronization of the unknown directed signed network, while ensuring uniform semiglobal practical asymptotic stability of the estimation errors. Numerical simulations validate the theoretical results.
\end{abstract}

% \textcolor{blue}{\begin{keywords}
% Topology identification, signed networks, synchronization, multi-agent systems, repelling Laplacian.
% \end{keywords}}

%%%%%%%%%%%%%%%%%%%%%%%%%%%%%%%%%%%%%%%%%%%%%%%%%%%%%%%%%%%%%%%%%%%%%%%%%%%%%%%%
\section{INTRODUCTION}
Many natural or human-made systems can be viewed as networks of agents that exchange local information and adjust their behavior accordingly, giving rise to a collective behavior that a single agent cannot achieve alone. Examples include multi-robot systems that coordinate to achieve complex tasks, such as transporting an object while holding a formation; and groups of individuals whose opinions evolve through social interactions \cite{newman2003structure}. A fundamental collective behavior in such systems is synchronization, where all agents converge to a common state or trajectory. Whether synchronization is achieved is governed by the interaction topology among the agents. In practice, however, this topology is often partially or entirely unknown and when interactions are one-directional, it need not be symmetric. 
% This motivates the problem of %simultaneously 
% identifying the interaction topology %and synchronizing the network 
% from the observed behavior.

%Dynamical networks are used to represent a large number of real networks, \textit{e.g.,} social networks, biological networks, multi-robot systems, sensor systems \cite{newman2003structure}. For instance, the human immune system can be viewed as a complex dynamical network where different types of cells communicate to defend the body against pathogens. Similarly, in multi-robot systems, autonomous agents exchange information and adjust their behaviors based on local interactions to achieve complex tasks. The collective behavior emerging from these local interactions can be modeled using complex dynamical networks. The interaction topology plays a key role in control design, particularly for achieving synchronization, yet it may be partially or entirely unknown in practice.

Existing approaches to topology identification include prediction error-based \cite{weerts2018prediction}, adaptive control-based \cite{zhu_new_2021}, constrained Lyapunov equation-based \cite{van_waarde_topology_2019}, and learning-based methods \cite{11045067}, among others. In this work, we focus on adaptive control-based topology identification methods, in which the network is controlled to track an auxiliary network. These methods guarantee correct identification under linear independence conditions, which are hard to maintain once the network synchronizes. To overcome this, \cite{zhu_new_2021, wang2023finite, restrepo2023simultaneousTCNS} introduce external inputs that prevent synchronization and provide persistency of excitation.
%In the literature on topology identification, existing approaches include prediction error-based \cite{weerts2018prediction}, adaptive control-based \cite{zhu_new_2021}, constrained Lyapunov equation-based \cite{van_waarde_topology_2019}, causality-based \cite{li2023topology} methods, and learning-based methods \cite{11045067}. In this work, we focus on adaptive control-based methods for topology identification, where the control of the original network is designed to track an auxiliary network. Under certain linear independence conditions, these methods guarantee correct topology identification. However, once the network achieves synchronization, such conditions become difficult to verify or maintain. To overcome this, works in \cite{zhu_new_2021, wang2023finite, restrepo2023simultaneous,restrepo2023simultaneousTCNS} introduce external inputs to prevent synchronization and provide the persistency of excitation for the topology estimation.

In all the above references, however, agents are assumed to cooperate with each other, whereas many scenarios involve antagonistic behaviors, for instance in robotics applications \cite{IEEE-robotmanip} and in social networks \cite{Fontan2021Signed}. A common framework for modeling dynamical systems with both cooperative and antagonistic interactions is that of \textit{signed graphs} \cite{altafini2012_6329411}, in which cooperative interactions correspond to positive edge weights and antagonistic interactions to negative ones. In this work, we model such systems using the \textit{repelling} signed Laplacian, whose main feature is that it always has a zero eigenvalue, but, since its diagonal entries can be negative, it may also have eigenvalues with negative real part, in which case the uncontrolled network is unstable. Conditions for the stability of the repelling Laplacian can be found in \cite{fontan2022multiagent}. The problem of simultaneous topology identification and synchronization was addressed in \cite{sekercioglu2026topology} for undirected signed graphs with weights of magnitude at most one.

%In all the references above, agents are assumed to cooperate with each other. However, many real-world scenarios involve antagonistic behaviors, such as in robotics applications \cite{csekerciouglu2024distributed} and social networks where agents compete \cite{Fontan2021Signed}. A common framework for modeling both cooperation and antagonism in dynamical systems is that of \textit{signed graphs} \cite{altafini2012_6329411,Shi2019Dynamics}. In the current literature, there are two different definitions of signed Laplacians for signed graphs, which depend on how the diagonal elements are calculated from the weighted adjacency matrix. They are called \textit{opposing} and \textit{repelling} signed Laplacians in \cite{Shi2019Dynamics}. In this paper, we focus on \textit{repelling} signed Laplacian, whose main feature is that it always has 0 as an eigenvalue, but may or may not be stable since it can have diagonal negative entries. Conditions for the stability of the repelling Laplacian can be found in \cite{9965602}.
% for networks interconnected oved directed signed graphs
%this to be rewritten when the technical part is over , where the agents are interconnected 
In this paper, we consider directed signed graphs with weights possibly greater than one, and address the problem of simultaneous topology identification and synchronization. Existing adaptive-control-based results cover either directed but all cooperative interactions \cite{restrepo2023simultaneousTCNS} or signed but undirected networks with normalized weights \cite{sekercioglu2026topology}: we treat the case that fuses both. We recast the problem through a \textit{node-centric regressor} and \textit{repelling signed Laplacians}. As in \cite{restrepo2023simultaneousTCNS,sekercioglu2026topology}, simultaneous synchronization and topology identification is cast as a stabilization problem of the origin in the error coordinates, using an auxiliary network driven by a $\delta$-persistently exciting input \cite{panteley2001relaxed}. Our contributions are threefold. First, we establish uniform global asymptotic stability in terms of the errors, even when the overall system is initially unstable, provided some global information about the network is available, \textit{e.g.,} the smallest eigenvalue of the symmetric part of the repelling signed Laplacian. Second, we relax this requirement by proposing a time-varying adaptive control gain. Third, we show uniform semi-global practical asymptotic stability of the origin in terms of the signed topology estimation errors when the excitation depends only on the measurable synchronization error.

\section{Preliminaries}
\textit{Notation:} $\lvert \cdot \rvert$ denotes the absolute value for scalars, the Euclidean norm for vectors, and the spectral norm for matrices. $\mbox{diag}(\bm z)$ denotes a diagonal matrix whose diagonal elements are the entries of the vector $\bm z$. $\operatorname{col}(z_i)$ denotes the column vector with entries $z_i$. $\mathbb{R}$ is the set of real numbers and $\mathbb{R}_{\geq 0}$ the nonnegative orthant. $A>0$ ($A\ge 0$) indicates that $A$ is a positive definite (positive semidefinite) matrix. 
%A continuous function $\alpha:\mathbb{R}_{\geq 0}\to \mathbb{R}_{\geq 0} \in \mathcal{K}$ if it is strictly increasing and $\alpha(0)=0$; $\alpha \in \mathcal{K}_{\infty}$ if, in addition, $\alpha(s)\to \infty$ as $s \to \infty$. 
We denote by $\mathbb{B}(\delta) \subset \mathbb{R}^n$ the open ball of radius $\delta$ centered at the origin, \textit{i.e.,} $\mathbb{B}(\delta) := \{ \bm x \in \mathbb{R}^n : \lvert \bm x \rvert < \delta\}$. We use the notation $\mathbb{H}(\delta,\Delta) := \{ \bm x \in \mathbb{R}^n : \delta \leq |\bm x| \leq \Delta\}$. 
%\rednotes{AR: We can define \(A_{\mathrm{sym}}\) here... and maybe make the vector bold.} \bluenotes{PS:Do we use $A_{\mathrm{sym}}$ in the paper? Then should we also define $L_{\mathrm{sym}}$? Otherwise, we don't here and just leave them in the rest of the paper. If you put vectors always bold in the paper, yes we can do that!}
%\rednotes{AR: What I meant was that we can define what \(sym\) means in the denominator.. and for my papers I always define vectors bold, otherwise it gets difficult to distinguish between vectors and scalars.. but I would like your input as well.}
%\bluenotes{PS: Ah ok, we can define sym! I never do that but I am not against it, I think it is easier to read as well. So, please do! :)}

\subsection{Signed graph theory}
\label{sec:prel_signed}
Let $\Gs = (\mathcal{V}, \mathcal{E})$ denote a signed graph, where $\mathcal{V} = \{ \nu_1, \nu_2, \dots, \nu_N \}$ is the set of nodes and $\mathcal{E} \subseteq \mathcal{V} \times \mathcal{V}$ is the set of $M$ edges, with self-loops not considered.  The adjacency matrix of \( \Gs \) is \( A := [a_{ij}] \in \mathbb{R}^{N \times N} \), where \( a_{ij} \neq 0 \) if and only if \( (\nu_j, \nu_i) \in \mathcal{E} \). Each edge $\varepsilon_k := (\nu_j, \nu_i)$ in $\mathcal{E}$ has a sign, either positive ($a_{ij} > 0$) representing a cooperative relationship or negative ($a_{ij} < 0$) representing an antagonistic relationship. If all edges have positive signs, then the graph is called an \textit{unsigned} graph. The graph is \textit{undirected} if the information flow between agents is bidirectional, meaning $ ( \nu_i, \nu_j )  \in \mathcal{E}$ $\Longleftrightarrow ( \nu_j, \nu_i ) \in \mathcal{E}$ with $a_{ij}=a_{ji}$ for all $i,j \leq N$. Otherwise, the graph is \textit{directed}. {The graph $\Gs$ is said to be \textit{complete} if $(\nu_j, \nu_i)\in \mathcal{E}$ for every pair $i \ne j$, that is, every ordered pair of distinct nodes is interconnected by an edge.} A rooted directed spanning tree of $\Gs$ is a subgraph that includes all the vertices of the original graph and contains a root node from which every other node is reachable via a directed path. 
%The incidence matrix $E \in \mathbb{R}^{N \times M}$ of $\Gs$ is defined by ${[E]_{ik}} :=+1,$ if $\nu_i$ is the initial node of the edge $\varepsilon_k$; ${[E]_{ik}} := -1,$ if $\nu_i$ is the terminal node of the edge $\varepsilon_k$; ${[E]_{ik}} := 0$, otherwise. The in-incidence matrix $E_{\odot} \in \mathbb{R}^{N \times M}$ of $\Gs$ is defined by ${[E_{\odot}]_{ik}} :=-1,$ if $\nu_i$ is the terminal node of the edge $\varepsilon_k$, and ${[E_{\odot}]_{ik}} := 0$, otherwise. The weight matrix $W_s := \mbox{diag}\{ w_{s_k} \} \in \mathbb{R}^{M \times M}$ contains the signed weights, where the weight of the edge $\varepsilon_k = (\nu_j, \nu_i)$ is $w_{s_k}=a_{ij}$. 
The repelling Laplacian $L \in \mathbb{R}^{N \times N}$ of a directed signed graph $\Gs$ is defined by 
%$L = E_{\odot}W_sE^\top$, {with entries given by 
$\ell_{ii} = \sum_{j \neq i} a_{ij}$ and $\ell_{ij} = -a_{ij}$ for $i \neq j$.

%\rednotes{AR: I meant comments like this on the PDF file. It is easy to check and correct this way. You can select a different color. I have made \textbackslash bluenotes, but you can make more.}
%\rednotes{AR: Also, I think the incidence matrix definition might not be needed as we are looking at the node centric formulation.}

\subsection{Persistency of excitation}
%\rednotes{PS: We need to rewrite a bit the connection sentences, not to have them exactly the same as in the 2026 ECC paper.}

%\rednotes{AR: The magenta colored text is revised form the older version (blue).}

%\bluenotes{We recall the following definitions from \cite{panteley2001relaxed} on ($\delta$-) persistency of excitation and lemmata from \cite{panteley2001relaxed,loria2005nested,loria1999new}, which serve as the basis for our stability analysis of the estimation errors. Consider the system }

In this subsection, we review the definitions of uniform and $\delta$-persistency of excitation \cite{panteley2001relaxed}, along with two stability lemmas \cite{panteley2001relaxed,loria1999new}. Consider the time-varying system
\begin{equation}\label{104}
    \dot{\bm  x} = f(t,\bm x),
\end{equation}
where $f:\mathbb{R}_{\ge0}\times\mathbb{R}^{n}\to\mathbb{R}^{n}$ with $n:=n_1+n_2$ such that all solutions $\bm x(t,t_0,\bm x_0)$ with $\bm x(0)=\bm x_0$ are defined for all $t\ge t_0$. Let $\bm x^\top := [\bm x_1^\top \ \bm x_2^\top]$, with $\bm x_1 \in \mathbb{R}^{n_1}$ and $\bm x_2 \in \mathbb{R}^{n_2}$.

\begin{definition} (\cite{panteley2001relaxed})\label{PE} Let $p,q\ge1$ and $\phi:\mathbb{R}_{\ge0}\times\mathbb{R}^{n}\to\mathbb{R}^{p\times q}$ be
%$\phi(\cdot)$ is 
bounded and globally Lipschitz. The pair $(\phi,f)$ is called \textit{uniformly persistently exciting} (\(u\)-PE) if, for each $r>0$, there exist $\mu, T>0$ such that for all $(t_0, \bm x_0) \in \mathbb{R}_{\ge0} \times \mathbb{B}(r)$, all solutions of \eqref{104} satisfy $\int_t^{t+T}\phi(\tau,\bm x(\tau,t_0,\bm x_0))\phi(\tau,\bm x(\tau,t_0,\bm x_0))^{\top}d\tau \ge \mu I, \forall t\geq t_0$.
\end{definition}
\begin{definition} (\cite{panteley2001relaxed})\label{delta-PE} The pair $(\phi,f)$ is called \textit{uniformly $\delta$-persistently exciting} ($u\delta$-PE) with respect to $\bm x_1$ if, for each $r$ and $\delta>0$, there exist constants $T(r,\delta)>0$ and $\mu(r,\delta)>0$, such that for all $(t_0, \bm x_0) \in \mathbb{R}_{\ge0} \times \mathbb{B}(r)$, all solutions of \eqref{104} satisfy $\min_{s\in [t, t+T]} |\bm x_1(s)| \geq \delta \Rightarrow$
$\int_t^{t+T}\phi(\tau,\bm x(\tau,t_0,\bm x_0))\phi(\tau,\bm x(\tau,t_0,\bm x_0))^{\top}d\tau \ge \mu I, \forall t\geq t_0$.
\end{definition}
% With an abuse of notation, we call the function $\phi$, $u\delta$-PE with respect to $x_1$ if the pair $(\phi,f)$ is $u\delta$-PE with respect to $x_1$. Define the set $\mathcal{D} := (\mathbb{R}^{n_1} \setminus \{ 0\}) \times \mathbb{R}^{n_2}$. Then, we have the following.% lemma from \cite{loria2005nested}.
% \begin{lemma}(\cite[Lemma 1]{loria2005nested})\label{lemma2} If $x \mapsto \phi(t,x)$ is continuous uniformly in $t$, then $\phi$ is $u\delta$-PE with respect to $x_1$ if and only if for each $x \in \mathcal{D}$, there exist $T>0$ and $\mu >0$ such that $\int_t^{t+T}|\phi(\tau,x)|d\tau \geq \mu$ for all $t \in \mathbb{R}$.
% \end{lemma}

%\bluenotes{The following lemma establishes that the output of a strictly proper, minimum phase and strictly stable filter, driven by a $u\delta$-PE input signal, is also $u\delta$-PE. A reminiscent version of the Lemma is originally presented in \cite{panteley2001relaxed}.}

The next lemma, adapted from \cite{panteley2001relaxed}, shows that passing a $u\delta$-PE input through a stable, strictly proper, and minimum-phase filter preserves the $u\delta$-PE property at the output.
\begin{lemma} \label{filprop}
    Let $\phi:\mathbb{R}_{\ge0}\times\mathbb{R}^{n_1}\to\mathbb{R}^{n_2}$ and consider
\begin{align}\label{eq:filtration_system}
\begin{bmatrix}
\dot{\bm x}\\
\dot{\bm \omega}
\end{bmatrix}
=
\begin{bmatrix}
g(t, \bm x, \bm \omega) \\
f_1(t, \bm \omega) + f_2(t, \bm x)\bm \omega + \phi(t, \bm x)
\end{bmatrix}
\end{align}
with $g:\mathbb{R}_{\ge0}\times\mathbb{R}^{n_1}\times\mathbb{R}^{n_2}\to\mathbb{R}^{n_1}$, $f_1:\mathbb{R}_{\ge0}\times\mathbb{R}^{n_2}\to\mathbb{R}^{n_2}$ Lipschitz in $\bm \omega$ uniformly in $t$ and measurable in $t$, and satisfying $|f_1(\cdot)| \le |\bm \omega|$ for all $t$;
$f_2:\mathbb{R}_{\ge0}\times\mathbb{R}^{n_1}\to\mathbb{R}^{n_2\times n_2}$ is locally Lipschitz in $\bm x$ uniformly in $t$ and measurable in $t$.
Assume that $\phi(t, \bm x)$ is $u\delta$-PE with respect to $\bm x$. Assume also that $\phi$ is locally Lipschitz and there exists a non-decreasing function $\alpha : \mathbb{R}_{\ge 0} \to \mathbb{R}_{\ge 0}$ such that, for almost all $(t,\bm x)\in\mathbb{R}_{\ge0}\times\mathbb{R}^{n_1}$:
\begin{equation}
\max \left\{
|\phi(\cdot)|,\, |g(\cdot)|,\, |f_2(\cdot)|,\,
\left| \frac{\partial \phi(\cdot)}{\partial t} \right|,\,
\left| \frac{\partial \phi(\cdot)}{\partial \bm x} \right|
\right\}
\le \alpha(|\bm x|).
\label{eq:filtration_alpha}
\end{equation}
Assume further that all solutions $t \mapsto \bm x_\phi$ of \eqref{eq:filtration_system}, with $\bm x_\phi^{\top} = [\bm x^{\top} \; \bm \omega^{\top}]$, are defined in $[t_0, \infty)$ and satisfy
$|\bm x_\phi(t, t_0, \bm x_{\phi_0})| \le r, \forall t \ge t_0,$ then $\bm \omega$ is $u\delta$-PE with respect to $\bm x$.
\end{lemma}
%\bluenotes{Finally, the following lemma from \cite{loria1999new} will be used in the subsequent analysis.}

Lastly, we state a stability result from \cite{loria1999new} that will be used to prove the convergence of the closed-loop error system.
%our main results on the stability analysis (in Propositions \ref{prop1} and \ref{prop2}).
\begin{lemma} \label{kyplemma}
For the system 
\begin{align}\label{lemma2eq}
\begin{bmatrix}
		\dot{{\bm x}}_1 \\ \dot{\bm x}_2 
	\end{bmatrix} = 
	\begin{bmatrix}
		A&
			B \psi(t,\bm x(t))^\top \\
			-\psi(t,\bm x(t))C^\top & 0
	\end{bmatrix} 	\begin{bmatrix}
		\bm x_1 \\ \bm x_2
	\end{bmatrix}\end{align}
suppose that there exist continuous non-decreasing functions 
$\alpha_i : \mathbb{R}_{\ge 0} \to \mathbb{R}_{\ge 0}$, $i = 1, 2$, such that $|\psi(t, \bm x)| \le \alpha_1(|\bm x|),$
$$\max \left\{
\left| \frac{\partial \psi(t, \bm x)}{\partial t} \right|,
\left| \frac{\partial \psi(t, \bm x)}{\partial \bm x} \right|
\right\} \le \alpha_2(|\bm x|),$$
and assume that $\psi(t, \bm x(t))$ is $u\delta$-PE with respect to $\bm x$.
If the triple $(A, B, C)$ satisfies the Kalman--Yakubovich--Popov (KYP) lemma, i.e., there exist $P = P^{\top} > 0$ and $Q = Q^{\top} > 0$ such that
$A^{\top}P + PA = -Q$ and $PB = C$, then the origin of \eqref{lemma2eq} is uniformly globally
asymptotically stable.
\end{lemma}

\section{Model and Problem Formulation}\label{sec:m-p-formulation}

%\rednotes{AR: We can add disturbance, but then we can not guarantee that the values of the estimated weights converge to the true weights. The error system instead of being uniformly globally asymptotically stable... it becomes uniformly globally bounded in Proposition 1. practically, for small disturbances, we notice noise in the estimated weights.. which can be removed by getting a robust term in the input. Let me know if you want to handle it, it changes quite a bit of the paper, and it will take until end of this week, maybe... The main issue is that we converge to the neighborhood of the weights.. but not the exact weights...}

Consider a network of $N$ agents exchanging information over an unknown directed signed graph to update their states. Each agent $i \in \{1, \dots, N\}$ updates its state according to the following dynamics
\begin{equation}
    \dot{x}_i =  - \sum_{j \neq i} a_{ij}(x_i - x_j) + u_i, 
    \label{eq:node_in_a_system_equation}
\end{equation}
%\rednotes{AR: Do we want to remove \(f_i(x_i)\); it is redundant; other way would be to bound it, by Lipschitz or other properties like One-sided lipschitz. and update the controller, but that will be more work.} \bluenotes{PS: Let's not do more work I think and just remove it. We just need to make sure that we remove it from everywhere ;)}
%$f_i(x_i)$ captures the intrinsic nonlinear dynamics,
where $x_i \in \mathbb{R}$ represents the state of agent $i$,  and $u_i$ is the external control input. The scalar $a_{ij} \in \mathbb{R}$ quantifies the unknown directed influence that agent $j$ exerts on agent $i$, which can either be {a} cooperative ($a_{ij}>0$) or an antagonistic ($a_{ij}<0$) interaction.

By defining the aggregated state vector $\bm x = [x_1, \dots, x_N]^\top$,  and the control input $\bm u = [u_1, \dots, u_N]^\top$, the network dynamics can be compactly expressed as %the intrinsic dynamics $F(\bm x) = [f_1(x_1), \dots, f_N(x_N)]^\top$,
\begin{equation}
    \dot{\bm x} = - L\bm x + \bm u, %F(\bm x) 
    \label{eq:systemequation}
\end{equation}
% Replace Equation (6)
%\begin{equation}
%   \magentanotes{ \dot{\bm x} = F(\bm x) - L\bm x + \bm u + \bm d(t),}
%    \label{eq:systemequation}
%\end{equation}
where $L \in \mathbb{R}^{N \times N}$ is the repelling signed Laplacian matrix defined in Section~\ref{sec:prel_signed}. % analogously to its unsigned counterpart, with entries  $\ell_{ii} = \sum_{j \neq i} a_{ij}$ and $\ell_{ij} = -a_{ij}$ for $i \neq j$.
By definition, $0$ is always an eigenvalue of $L$, with the vector of ones $\mathds{1} \in \mathbb{R}^N$ as the corresponding right eigenvector \cite{fontan2022multiagent}. In contrast to the unsigned Laplacian, the repelling signed Laplacian can be indefinite---possessing both positive and negative eigenvalues---or have multiple zero eigenvalues. Consequently, the uncontrolled system \eqref{eq:systemequation} can exhibit diverse behaviors, including consensus, clustering, or unstable divergent dynamics \cite{fontan2022multiagent}. %\rednotes{AR: Maybe adding the citation helps. I dont want to define clustering in the middle.. maybe we write some literature review in the introduction.. and explain a little in it...}. \bluenotes{PS: yes I think it is a good idea.} 
Furthermore, due to the directed nature of the graph, $L$ is asymmetric ($L \neq L^\top$). {Therefore, the standard factorization utilized for undirected signed topologies, which relies on the incidence and weight matrices, is no longer applicable (see \cite{restrepo2023simultaneousTCNS,sekercioglu2026topology} for more details on the undirected factorization).}

To address this asymmetry and the fact that the true network topology is unknown, we approach the directed topology identification problem by considering a complete graph encompassing all $N(N-1)$ possible directed edges. %We define a global stacked parameter vector $\bm{w} \in \mathbb{R}^{N(N-1)}$ containing these potential weights. Each scalar entry of this vector corresponds to a specific directional interaction $a_{ij}$ (for $i,j \in \{1, \dots, N\}, i \neq j$), such that $a_{ij} \neq 0$ if the edge exists in the true graph, and $a_{ij} = 0$ otherwise.
We define a global stacked parameter vector $\bm{w} = \operatorname{col}(w_k) \in \mathbb{R}^{N(N-1)}$ ($k \in \{1, \dots, N(N-1)\}$) containing these potential weights. Each scalar entry $w_k$ of this vector corresponds to a specific directional interaction $a_{ij}$ (for $i,j \in \{1, \dots, N\}, i \neq j$), such that $a_{ij} \neq 0$ if the edge exists in the true graph, and $a_{ij} = 0$ otherwise.

%To address this asymmetry and the fact that the true network topology is unknown, we approach the directed topology identification problem by considering a complete graph encompassing all $N(N-1)$ possible directed edges. We define a stacked parameter vector $\bm w \in \mathbb{R}^{N(N-1)}$ containing these potential weights, such that an entry $w_k \neq 0$ if the corresponding interaction exists in the true graph, and $w_k = 0$ otherwise. \rednotes{PS: Here we should add $k \in \{ 1, 2 , \dots, N(N-1)\}$ (or until $(N-1)$?) Also, it is a bit confusing to have $w_k$ here but $w_i$ below. So maybe we keep $w_k$ since $i$ and $j$ are $\leq N$.}

To isolate these unknown directed weights from the state variables, we utilize a node-centric regressor formulation. For each agent $i \in \{1, \dots, N\}$, let $\bm{w}_i := \operatorname{col}(a_{ij})_{j \ne i} \in \mathbb{R}^{N-1}$ be the local sub-vector comprising all potential incoming edge weights to node $i$. Correspondingly, let $\bm{x}_{\Delta i} := \operatorname{col}(x_i-x_j)_{j \ne i} \in \mathbb{R}^{N-1}$ be the vector of associated state differences. This allows the Laplacian operation to be factored into a block-diagonal regressor matrix $\Psi(\bm{x}) \in \mathbb{R}^{N \times N(N-1)}$ and the global parameter vector $\bm{w} := \operatorname{col}(\bm{w}_1, \dots, \bm{w}_N)$ as:

%To isolate these unknown directed weights from the state variables, we utilize a node-centric regressor formulation. For each agent $i$ \rednotes{PS: to remind the readers $i \leq N$ and also for $j$ I guess to: $i,j \leq N$.}, let $\bm w_i=\operatorname{col}(a_{ij})_{j\ne i} \in \mathbb{R}^{N-1}$  be the vector comprising all potential incoming edge weights to node $i$, \rednotes{PS: here as I said before, maybe we use $w_k$ and write $k \leq ...$} and $\bm x_{\Delta i}:=\operatorname{col}(x_i-x_j)_{j\ne i}. \in \mathbb{R}^{N-1}$ be the vector of corresponding state differences. \rednotes{PS: Same for $x_{\Delta i}$, is it until $N$ or $N-1$? If not N then we should use $k$ or some other symbol.} This allows the Laplacian operation to be factored into a block-diagonal regressor matrix $\Psi(\bm x) \in \mathbb{R}^{N \times N(N-1)}$ and the stacked parameter vector $\bm w$ as

\begin{equation}
    L\bm x = \Psi(\bm x)\bm w,
\end{equation}
where $\Psi(\bm x) = \text{blkdiag}(\bm x_{\Delta 1}^\top, \bm x_{\Delta 2}^\top, \dots, \bm x_{\Delta N}^\top)$.

The  control objective of this work is to design the external control protocol $\bm u$ to achieve simultaneous network synchronization and topology identification. We say that the network synchronizes if all agents converge to a common state, i.e., $\lim_{t\to\infty}(x_i(t)-x_j(t))=0$ for all $i,j$. A successful identification is achieved when an estimate of the parameters, denoted as $\hat{\bm w}$, converges to the true parameter vector $\bm w$ (i.e., $\lim_{t \to \infty} (\bm w - \hat{\bm w}(t)) = 0$), revealing the exact presence, direction, and sign of every network interaction. %In the following section, we introduce an adaptive control strategy and an auxiliary network designed to be PE to fulfill this dual objective.
%\rednotes{AR: Not sure what you meant by adding a small figure!!} \bluenotes{PS: Not sure we'll have space finally but I can send you an example of a figure by email, it was just an idea.} \rednotes{AR: We have 8 pages in total!, We can remove it for the final revision to make it 6 pages. or we remove the proofs...} \bluenotes{PS: I think what we will have to do is to put the equations without numbering in line, even if it looks bad :D.}

%To address this asymmetry and isolate the unknown directed weights, we utilize a node centric regressor formulation. For each agent $i$, let $w_i:=\operatorname{col}(a_{ij})_{j\ne i} \in \mathbb{R}^{N-1}$ be the vector comprising all incoming edge weights to node $i$, and $x_{\Delta i}:=\operatorname{col}(x_i-x_j)_{j\ne i}. \in \mathbb{R}^{N-1}$ be the vector of corresponding state differences. This allows the Laplacian operation to be factored into a block-diagonal regressor matrix $\Psi(x) \in \mathbb{R}^{N \times N(N-1)}$ and a stacked parameter vector $w \in \mathbb{R}^{N(N-1)}$
\section{Main Results}
%\rednotes{PS: I think what is missing here is a short paragraph to first explain what we do. Something like: "we design an auxiliary variable $\hat x$ that is PE (See Def 1) and ensure our network (5 / 6) converges to is in order to achieve synchronization and identification. Then, we redesign the control gain in terms of the synchronization errors, and finally we show uspas etc."}

{In this section, we formulate the proposed framework for simultaneous network synchronization and topology identification over directed signed networks. First, we introduce an auxiliary filter $\hat{\bm x}(t)$ driven by a persistently exciting (PE) signal (Definition \ref{PE}) to inject the required excitation for exact parameter estimation. Next, we design an adaptive control gain dependent only on local synchronization errors to eliminate the reliance on global network knowledge. Finally, we establish uniform semi-global practical asymptotic stability of the closed-loop system for practical implementations where the excitation signal depends exclusively on measurable tracking errors.}

To this end, we define the auxiliary system $\hat{\bm x}(t) \in \mathbb{R}^N$ driven by the PE input $\phi(t, \bm \xi)$:
\begin{equation}
    \dot{\hat{\bm x}}(t) = -\hat{\bm x}(t) + \phi(t, \bm \xi),
    \label{eq:auxiliary_sys}
\end{equation}
where $\bm \xi^\top := [\tilde{\bm x}^\top \ \tilde{\bm w}^\top]$ represents the augmented error coordinates (defined below), and  $\phi(t, \bm \xi): \mathbb{R}_{\geq 0} \times \mathbb{R}^{N+ N(N-1)} \to \mathbb{R}^{N}$, $(t,\bm \xi)\mapsto \phi(t,\bm \xi)$ is a function satisfying the following assumption and its excitation properties are specified in Definitions \ref{PE}-\ref{delta-PE}. %(Correct me if I am wrong with the dimensions please.)} satisfies the following assumption. $\phi(t, \bm \xi)$ \rednotes{PS:
%\rednotes{PS: I was wondering, only if it is easy, if we could add something like $\hat L \hat x$ in Eq. (8), to make the network synchronize with respect to the interaction graph and not necessarily to zero. But then I imagine the problem would be if the graph is initially unstable, $\hat L$ is also unstable so it will make the system diverge... so maybe not a great idea...}

\begin{ass}\label{assumption1}
Let $\rho:\mathbb{R}_{\geq 0}\to \mathbb{R}_{\geq 0}$ be a continuous non-decreasing function. For all $\bm \xi \in \mathbb{R}^{N+N(N-1)}$ and almost all $t \in \mathbb{R}_{\geq 0}$, $\phi(t, \bm \xi)$ satisfies:
\begin{align*}
    \max \left\{ \lvert \phi(\cdot) \rvert, \lvert \frac{\partial \phi(\cdot)}{\partial t} \rvert , \lvert \frac{\partial \phi(\cdot)}{\partial \bm \xi} \rvert \right\} \leq \rho(\lvert \bm \xi \rvert).
\end{align*}
\end{ass}

%\begin{ass}\label{ass:disturbance}
%\magentanotes{The external disturbance vector $\bm d(t) \in \mathbb{R}^N$ is piecewise continuous and uniformly bounded such that $\sup_{t \ge 0} \lvert \bm d(t) \rvert \le \bar{d}$ (and component-wise $\sup_{t \ge 0} \lvert d_i(t) \rvert \le \bar{d}_\infty$), where $\bar{d}, \bar{d}_\infty > 0$ are positive constants.}
%\end{ass}

The proposed control protocol utilizes exact cancellation of the intrinsic dynamics while tracking the auxiliary system:
\begin{equation}
    \bm u =- c_1(\bm x - \hat{\bm x}(t)) + \dot{\hat{\bm x}}(t) + \hat{L}\hat{\bm x}(t), % -F(\bm x) 
\label{eq:controlinput}
\end{equation}
% Replace Equation (10)
%\begin{equation}
% \magentanotes{   \bm u = -F(\bm x) - c_1(\bm x - \hat{\bm x}(t)) + \dot{\hat{\bm x}}(t) + \hat{L}\hat{\bm x}(t) + \bm u_{\mathrm{rob}}(\tilde{\bm x}),}
%\label{eq:controlinput}
%\end{equation}
%\magentanotes{If an upper bound $k_d \ge \bar{d}_\infty$ is known, define:$$\bm u_{\mathrm{rob}}(\tilde{\bm x}) = -k_d \tanh(\beta \tilde{\bm x}),$$
%where $\tanh(\beta \tilde{\bm x}) := [\tanh(\beta \tilde{x}_1), \dots, \tanh(\beta \tilde{x}_N)]^\top$ and $\beta > 0$ is a steepness parameter. }
where $c_1 > 0$ is a scalar gain that needs to be designed and $\hat{L}\hat{\bm x}(t) = \Psi(\hat{\bm x}(t))\hat{\bm w}$ represents the estimated directed Laplacian operation based on the node-centric regressor.

To analyze the stability of the proposed protocol, we define the state synchronization error $\tilde{\bm x} := \bm x - \hat{\bm x}(t)$ and the parameter estimation error $\tilde{\bm w} := \bm w - \hat{\bm w}$. 
Evaluating the time derivative $\dot{\tilde{\bm x}} = \dot{\bm x} - \dot{\hat{\bm x}}$ and substituting \eqref{eq:systemequation} and \eqref{eq:controlinput} yields $\dot{\tilde{\bm x}} = -L\bm x - c_1\tilde{\bm x} + \Psi(\hat{\bm x}(t))\hat{\bm w}$. 
By using the equation $\bm x = \tilde{\bm x} + \hat{\bm x}$ and \((L\hat{x})_i = \sum_{j \neq i} {a_{ij}} {(\hat{x}_i - \hat{x}_j)} = \hat{x}_{\Delta i}^\top w_i\) which implies the regressor property $L\hat{\bm x}(t) = \Psi(\hat{\bm x}(t))\bm w$, we obtain the state error dynamics:

%\rednotes{$$(L\hat{x})_i = \sum_{j \neq i} \underbrace{a_{ij}}_{\text{in } w_i} \underbrace{(\hat{x}_i - \hat{x}_j)}_{\text{in } \Psi(\hat{x})} = \hat{x}_{\Delta i}^\top w_i \implies L\hat{x} = \Psi(\hat{x})w$$}

%\rednotes{$$(\hat{L}\hat{x})_i = \sum_{j \neq i} \underbrace{\hat{a}_{ij}}_{\text{in } \hat{w}_i} \underbrace{(\hat{x}_i - \hat{x}_j)}_{\text{in } \Psi(\hat{x})} = \hat{x}_{\Delta i}^\top \hat{w}_i \implies \hat{L}\hat{x} = \Psi(\hat{x})\hat{w}$$}

\begin{equation}
\dot{\tilde{\bm x}} = -(c_1 I + L)\tilde{\bm x} - \Psi(\hat{\bm x}(t))\tilde{\bm w}.
    \label{eq:error_dyn}
\end{equation}

% Replace Equation (11)
%\begin{equation}
%    \magentanotes{\dot{\tilde{\bm x}} = -(c_1 I + L)\tilde{\bm x} - \Psi(\hat{\bm x}(t))\tilde{\bm w} + \bm u_{\mathrm{rob}}(\tilde{\bm x}) + \bm d(t).}
%    \label{eq:error_dyn}
%\end{equation}

%To achieve the dual objectives of topology identification and state synchronization, we initially consider the scenario where a lower bound on the eigenvalues of the symmetric part of the directed Laplacian, defined as $L_{\mathrm{sym}} := \frac{1}{2}(L + L^\top)$, is known (i.e., \(\lambda_{\mathrm min}(L_{\mathrm sym})\)). \rednotes{PS: I was thinking, should the sentence on "initial scenario with lower bound using $L_{\mathrm{sym}}$" be moved to just before introducing Proposition 1? Since here, it is good to mention but then in Eq. (12), we don't really use it.} \rednotes{AR: Yes, I noticed it as well.. maybe merge it with the sentence just before prop. 1} We propose the following parameter update law:

To achieve the dual objectives of topology identification and state synchronization, we propose the following parameter update law
\begin{equation}
    \dot{\hat{\bm w}} = -\Psi(\hat{\bm x}(t))^\top \tilde{\bm x}.
    \label{eq:update_law}
\end{equation}

Noting that $\dot{\tilde{\bm w}} = -\dot{\hat{\bm w}}$ since the true parameter vector $\bm w$ is constant, the closed-loop error dynamics can be expressed in vector form as:
\begin{align}\label{eq:closed_loop}
	\begin{bmatrix}
		\dot{\tilde{\bm x}} \\ \dot{\tilde{\bm w}} 
	\end{bmatrix} = 
	\begin{bmatrix}
		-(c_1 I + L) & -\Psi(\hat{\bm x}(t)) \\
		\Psi(\hat{\bm x}(t))^\top & \mathbf{0}
	\end{bmatrix} 	\begin{bmatrix}
		\tilde{\bm x} \\ \tilde{\bm w}
	\end{bmatrix}.
\end{align}

\begin{comment}
    
\magentanotes{
% Replace Equation (13)
\begin{align}\label{eq:closed_loop}
	\begin{bmatrix}
		\dot{\tilde{\bm x}} \\ \dot{\tilde{\bm w}} 
	\end{bmatrix} = 
	\begin{bmatrix}
		-(c_1 I + L) & -\Psi(\hat{\bm x}(t)) \\
		\Psi(\hat{\bm x}(t))^\top & 0
	\end{bmatrix} 	\begin{bmatrix}
		\tilde{\bm x} \\ \tilde{\bm w}
	\end{bmatrix} + 
	\begin{bmatrix}
		\bm u_{\mathrm{rob}}(\tilde{\bm x}) + \bm d(t) \\
		0
	\end{bmatrix}.
\end{align}}
\end{comment}

%\rednotes{Need something here to introduce the proposition!!!}

In the literature on adaptive topology identification (see, e.g., \cite{zhu_new_2021}), error systems of the form \eqref{eq:closed_loop} are known to achieve global exponential stability when the auxiliary state $\hat{\bm x}(t)$ is designed to be PE. 
However, while a standard PE condition guarantees successful parameter estimation, its non-vanishing nature inherently prevents the network from achieving state synchronization. 
To overcome this limitation, recent works on undirected signed networks \cite{sekercioglu2026topology} and unsigned networks \cite{restrepo2023simultaneousTCNS} have demonstrated that simultaneous topology identification and synchronization can be achieved by utilizing a $u\delta$-PE signal (See Definition \ref{delta-PE}). 
Building upon this established framework, the following proposition extends the results of \cite{sekercioglu2026topology} from undirected to directed signed networks. We show that if a lower bound on the eigenvalues of the symmetric Laplacian, defined as $L_{\mathrm{sym}} := \frac{1}{2}(L + L^\top)$, is known and utilized to appropriately bound the control gain, designing the auxiliary input $\phi(t, \bm \xi)$ to be $u\delta$-PE guarantees that both control objectives are simultaneously met.
%Building upon this established framework, the following proposition extends the results of \cite{sekercioglu2026topology} from undirected to directed signed networks, demonstrating that designing the auxiliary input $\phi(t, \bm \xi)$ to be $u\delta$-PE ensures both control objectives are simultaneously met.

\begin{proposition}\label{prop1}
Consider the network dynamics \eqref{eq:systemequation} with the control input \eqref{eq:controlinput} and the parameter update law \eqref{eq:update_law}. Let $\phi: \mathbb{R}_{\geq 0} \times \mathbb{R}^{N+N(N-1)} \to \mathbb{R}^N$ satisfy Assumption \ref{assumption1}, and $\lambda_{\min}(L_{\mathrm{sym}})$ denote the smallest eigenvalue of the symmetric Laplacian $L_{\mathrm{sym}} = \frac{1}{2}(L+L^\top)$. If $\phi$ is $u\delta$-PE, and the control gain satisfies $c_1 > -\lambda_{\min}(L_{\mathrm{sym}})$, then the origin of the closed-loop error system \eqref{eq:closed_loop} is uniformly globally asymptotically stable. Consequently, the network achieves synchronization ($\lim_{t \to \infty} \tilde{\bm x}(t) = 0$) and the directed topology is identified ($\lim_{t \to \infty} \hat{\bm w}(t) = \bm w$).
\end{proposition}

\begin{proof}
We analyze the stability of the closed-loop error system \eqref{eq:closed_loop} by invoking Lemma \ref{kyplemma}. Let $\bm x_1 = \tilde{\bm x}$, $\bm x_2 = \tilde{\bm w}$, $A = -(c_1 I + L)$, and define the PE signal as $\psi(t, \bm x(t)) = \Psi(\hat{\bm x}(t))^\top$. By setting the matrices $B = -I$ and $C = -I$, the system \eqref{eq:closed_loop} matches the structure of \eqref{lemma2eq}.

To satisfy the KYP conditions of Lemma \ref{kyplemma}, there must exist symmetric positive definite matrices $P > 0$ and $Q > 0$ such that $PB = C$ and $A^\top P + PA = -Q$.
Choosing $P = I > 0$ satisfies $PB = C$, since $I(-I) = -I$. Evaluating the Lyapunov equation for $Q$ yields:
\begin{align*}
    A^\top P + PA &= -(c_1 I + L)^\top - (c_1 I + L) \nonumber \\
    &= -2c_1 I - L^\top - L \nonumber \\
    &= -2(c_1 I + L_{\mathrm{sym}}) = -Q. %\label{eq:kyp_Q}
\end{align*}

%\rednotes{AR: Maybe the comparisons should be written after the proof. It would look something like this in magenta... and remove the blue paragraph...}

%{By the assumption of the proposition, the control gain is chosen such that $c_1 > -\lambda_{\min}(L_{\mathrm{sym}})$. Consequently, the matrix $(c_1 I + L_{\mathrm{sym}})$ is strictly positive definite. Therefore, $Q = 2(c_1 I + L_{\mathrm{sym}}) > 0$, fulfilling all algebraic conditions of the KYP lemma.}

By the assumption of the proposition, the control gain is chosen such that $c_1 > -\lambda_{\min}(L_{\mathrm{sym}})$. Consequently, the matrix $(c_1 I + L_{\mathrm{sym}})$ is strictly positive definite, and $Q = 2(c_1 I + L_{\mathrm{sym}}) > 0$, fulfilling all algebraic conditions of the KYP lemma. Moreover, considering the Lyapunov function $V(\bm \xi) = \frac{1}{2}\bm \xi^\top P \bm \xi = \frac{1}{2}\lvert\tilde{\bm x}\rvert^2 + \frac{1}{2}\lvert\tilde{\bm w}\rvert^2$, its time derivative along the trajectories of \eqref{eq:closed_loop} satisfies $\dot{V}(\bm \xi) = -\frac{1}{2}\tilde{\bm x}^\top Q \tilde{\bm x} \le 0$. This implies that the system \eqref{eq:closed_loop} is uniformly globally stable and its solutions $\bm \xi(t)$ are uniformly globally bounded.

Since $\bm \xi(t)$ is uniformly globally bounded, Assumption \ref{assumption1} ensures that the excitation signal $\phi(t, \bm \xi)$ is also uniformly bounded. For the auxiliary system \eqref{eq:auxiliary_sys}, consider the candidate Lyapunov function $\hat{V}(\hat{\bm x}) = \frac{1}{2}\lvert\hat{\bm x}\rvert^2$, whose time derivative satisfies
%For the auxiliary system \eqref{eq:auxiliary_sys}, consider the candidate Lyapunov function $\hat{V}(\hat{\bm x}) = \frac{1}{2}\lvert\hat{\bm x}\rvert^2$. Its derivative satisfies 
$$\dot{\hat{V}}(\hat{\bm x}) = -\lvert\hat{\bm x}\rvert^2 + \hat{\bm x}^\top\phi(t, \bm \xi).$$ 
After applying Young's inequality ($ab \le \frac{1}{2}a^2 + \frac{1}{2}b^2$), we get
\begin{equation*}
    \hat{\bm x}^\top\phi(t, \bm \xi) \le \frac{1}{2}\lvert\hat{\bm x}\rvert^2 + \frac{1}{2}\lvert\phi(t, \bm \xi)\rvert^2.
\end{equation*}
which, upon substitution into the Lyapunov derivative, gives
\begin{equation*}
    \dot{\hat{V}}(\hat{\bm x}) \le -\frac{1}{2}\lvert\hat{\bm x}\rvert^2 + \frac{1}{2}\lvert\phi(t, \bm \xi)\rvert^2 = -\hat{V}(\hat{\bm x}) + \frac{1}{2}\lvert\phi(t, \bm \xi)\rvert^2.
\end{equation*}
Consequently, whenever $\hat{V}(\hat{\bm x}) > \frac{1}{2}\lvert\phi(t, \bm \xi)\rvert^2$, the derivative $\dot{\hat{V}}(\hat{\bm x})$ becomes strictly negative. This confirms that the auxiliary system \eqref{eq:auxiliary_sys} is Input-to-State Stable with respect to the bounded input $\phi(t, \bm \xi)$, guaranteeing that $\hat{\bm x}(t)$ remains uniformly globally bounded \cite{KHALIL92}.

Because both $\bm \xi(t)$ and $\hat{\bm x}(t)$ are uniformly globally bounded and all conditions of Lemma \ref{filprop} are satisfied, the filtered state $\hat{\bm x}(t)$, and consequently the regressor $\Psi(\hat{\bm x}(t))$, is $u\delta$-PE with respect to $\tilde{\bm x}$. With the KYP conditions met and the regressor established as $u\delta$-PE, it follows from Lemma \ref{kyplemma} that the origin of \eqref{eq:closed_loop} is uniformly globally asymptotically stable.
\end{proof}

{Recall that, unlike standard unsigned networks, the Laplacian $L$ of a directed signed graph may possess eigenvalues with negative real parts, which consequently forces the symmetric part $L_{\mathrm{sym}}$ to have negative eigenvalues. The lower bound condition $c_1 > -\lambda_{\min}(L_{\mathrm{sym}})$ established in Proposition \ref{prop1} ensures that the control gain introduces sufficient damping to overcome any such repelling dynamics. Conversely, whenever the network topology satisfies $\lambda_{\min}(L_{\mathrm{sym}}) = 0$---such as in undirected or weight-balanced unsigned networks, or in undirected stable signed networks---any strictly positive gain $c_1 > 0$ inherently satisfies this requirement. }

{However, }Proposition \ref{prop1} relies on the exact knowledge of a global network parameter, specifically the smallest eigenvalue of the symmetric part of the directed Laplacian, $\lambda_{\min}(L_{\mathrm{sym}})$. Because the repelling signed Laplacian can be indefinite, identifying this minimum eigenvalue requires prior knowledge of the entire network's topology---including the exact presence, direction, and sign of all interactions.

%Proposition \ref{prop1} relies on the exact knowledge of a global network parameter, specifically the smallest eigenvalue of the symmetric part of the directed Laplacian, $\lambda_{\min}(L_{\mathrm{sym}})$. Because the repelling signed Laplacian can be indefinite, identifying this minimum eigenvalue requires prior knowledge of the entire network's topology---including the exact presence, direction, and sign of all interactions. 

%\rednotes{PS: In the following paragraph, you can also mention "Proposition 2" of [15] to be more specific.}
In the context of undirected signed networks, it was established in Proposition 2 of \cite{sekercioglu2026topology} that setting the control gain to $c_1 > N$ is sufficient to overcome the antagonistic interactions, provided the network features normalized weights $|a_{ij}| \leq 1$. However, while the $c_1 > N$ bound holds for undirected graphs, it is insufficient for directed signed graphs. Because directed repelling dynamics can compound asymmetrically within the symmetric Laplacian $L_{\mathrm{sym}}$, the minimum eigenvalue can drop well below $-N$.

%\rednotes{PS: I think this example is nice!
%The other thing, it may also be interesting to add a small remark/example for Proposition 1 as well, to show that choosing $c_1 >\lambda_{\min}(L)$ is not enough and what we get by $\lambda_{\min}(L_{sym})$. For instance L=[-1 1 0; 0 -1 1; 0 0 0] works I think. L has 0, -1 and -1 as eigenvalues but $L_{sym}$ has 0.24, -0.65 and -1.58 so just choosing $c > \lambda_{\min}(L)$ would not be enough. Maybe better to add it after Prop 1 or after the proof.}
%\bluenotes{AR: We can mention that $c_1 > -\lambda_{\min}(L_{\mathrm{sym}})$ is a sufficient condition, but not a necessary one. It might be that any $c_1 > -\lambda_{\min}(L)$ is also enough to achieve both synchronization and topology identification; for instance, in your example, $c_1 = 1.01$ actually works.} \rednotes{PS: Ah okok, then it is better not to give an example if it works in practice.}
\begin{example}
Consider a 3-agent directed network with
\[
    L=\begin{bmatrix}
        -2 &1& 1\\1 &0 &-1\\1&1&-2
    \end{bmatrix}.
\]
Evaluating the resulting symmetric Laplacian $L_{\mathrm{sym}}$ yields a minimum eigenvalue of $\lambda_{\min}(L_{\mathrm{sym}}) = -3.17$. In this case, applying the undirected stability bound of $c_1 > 3$ (e.g., $c_1 = 3.1$) fails to satisfy the strict requirement $c_1 > -\lambda_{\min}(L_{\mathrm{sym}})$ established in Proposition \ref{prop1}, and may consequently fail to stabilize the network. \qed
\end{example}

To relax this requirement and allow the control gain to be set without global structural knowledge, we present the following corollary for networks with normalized edge weights.

\begin{ass}\label{assumption2}
The unknown directed edge weights of the signed graph are normalized such that $|a_{ij}| \leq 1$ $\forall \,i,j$.
\end{ass}

\begin{corollary}\label{cor1}
Consider the closed-loop error system \eqref{eq:closed_loop} under the conditions of Proposition \ref{prop1} and let Assumption~\ref{assumption2} hold. If the auxiliary input $\phi(t, \bm \xi)$ is $u\delta$-PE, and the control gain is chosen such that $c_1 > 2(N-1)$, then the origin of the closed-loop error system \eqref{eq:closed_loop} is uniformly globally asymptotically stable, implying simultaneous synchronization and directed topology identification.
\end{corollary}
%\rednotes{PS: One suggestion: Maybe in the Corollary we just put $c_1 > 2(N-1)$ and not "the condition $c_1 > -\lambda_{\min}(L_{\mathrm{sym}})$ is satisfied" and then in the proof, from the beginning we say something like "To satisfy the stability requirement of Proposition \ref{prop1}, we must ensure $c_1 > -\lambda_{\min}(L_{\mathrm{sym}})$. Here, we show that $c_1 > 2(N-1)$ implies $c_1 > -\lambda_{\min}(L_{\mathrm{sym}})$ since $\lambda_{\min}(L_{\mathrm{sym}}) \geq -2(N-1)$". And then all the proof you did.}

\begin{proof}
To satisfy the stability requirement of Proposition \ref{prop1}, we must ensure $c_1 > -\lambda_{\min}(L_{\mathrm{sym}})$. Here, we show that choosing $c_1 > 2(N-1)$ satisfies this requirement by establishing that $\lambda_{\min}(L_{\mathrm{sym}}) \ge -2(N-1)$.

%To satisfy the stability requirement of Proposition \ref{prop1}, we must ensure $c_1 > -\lambda_{\min}(L_{\mathrm{sym}})$. %By the definition of the directed Laplacian $L$, its diagonal elements are $l_{ii} = \sum_{j \neq i} a_{ij}$ and its off-diagonal elements are $l_{ij} = -a_{ij}$ for $i \neq j$. 
The symmetric part of the Laplacian, $L_{\mathrm{sym}} $, possesses diagonal elements $ \sum_{j \neq i} a_{ij}$ and off-diagonal elements $ -\frac{1}{2}(a_{ij} + a_{ji})$. By Gershgorin circle Theorem \cite{horn2012matrix}, for at least one row $i \in \{1, \dots, N\}$, every eigenvalue $\lambda$ of $L_{\mathrm{sym}}$ must satisfy:
\begin{equation*}
    \lambda \ge  \sum_{j \neq i} a_{ij} - \frac{1}{2} \sum_{j \neq i} \left| a_{ij} + a_{ji} \right|.
\end{equation*}

%Under the assumption of normalized weights ($|a_{ij}| \le 1$), we evaluate the absolute minimum possible value for the terms inside the summation. For any pair of bounded weights $u, v \in [-1, 1]$, the function $f(u,v) = u - \frac{1}{2}|u+v|$ achieves its global minimum of $-2$ (occurring when $u = -1$ and $v = -1$). 

Under the assumption of normalized weights ($|a_{ij}| \le 1$), we evaluate the absolute minimum possible value for the terms inside the summation to find the leftmost boundary of any Gershgorin disk. For any pair of bounded weights $u, v \in [-1, 1]$, the function $f(u,v) = u - \frac{1}{2}|u+v|$ achieves its global minimum of $-2$ (occurring when both $u = -1$ and $v = -1$).

%\rednotes{PS: I think here it would be good to explain what we mean by "worst-case", meaning the graph is complete and every interconnection is antagonistic.} 
%\bluenotes{AR: Actually, the ``worst-case" here refers to the leftmost boundary of the Gershgorin disk for a single row, not necessarily the entire network topology. To push the Gershgorin bound to $-2(N-1)$, we only need a single node to have fully antagonistic bidirectional connections with all other $N-1$ nodes; the rest of the graph can be anything.} \rednotes{PS: Oh, Ok! That is interesting. So then, I think it would be interesting to make it clear. You maybe already did.}
Applying this minimum to all $N-1$ possible connections in the summation yields the lowest possible Gershgorin lower bound for any row, giving a strict lower bound for the spectrum:
%Applying this minimum bound to all $N-1$ off-diagonal elements in the worst-case row yields a strict lower bound for the spectrum:
\begin{equation*}
    \lambda_{\min}(L_{\mathrm{sym}}) \ge \sum_{j \neq i} (-2) = -2(N-1).
\end{equation*}
Therefore, setting the scalar gain $c_1 > 2(N-1)$  guarantees that the condition $c_1 > -\lambda_{\min}(L_{\mathrm{sym}})$ is satisfied, fulfilling all algebraic requirements of the KYP Lemma \ref{kyplemma} regardless of the unknown directed topology. The remainder of the proof follows directly from the proof of Proposition \ref{prop1}.
%Therefore, setting the scalar gain $c_1 > 2(N-1)$ unconditionally guarantees that the condition $c_1 > -\lambda_{\min}(L_{\mathrm{sym}})$ is satisfied, fulfilling all algebraic requirements of the KYP lemma regardless of the unknown sparse topology. \rednotes{PS: We could add a sentence like: "the remainder of the proof follows from the proof of Prop 1". Also, I think we should refer to Lemma 3 when you mention KYP lemma.}
\end{proof}

If the bounds of the directed weights and the minimum eigenvalue $\lambda_{\min}(L_{\mathrm{sym}})$ are entirely unknown, the system can be stabilized by implementing a time-varying adaptive control gain $c_1(t)$ governed by the update law:
\begin{equation}
    \dot{c}_1(t) = \gamma_c \tilde{\bm x}^\top \tilde{\bm x}
    \label{eq:adaptiveupdatelaw}
\end{equation}
where $\gamma_c > 0$ is a learning rate. {This approach aligns with adaptive methodologies that dynamically tune coupling strengths to avoid requiring global network information. For example, control gains can be designed to grow monotonically with consensus errors \cite{li2014designing} or increase slowly over time \cite{LCSS-Maitryee}. Similarly, the proposed adaptation law \eqref{eq:adaptiveupdatelaw} leverages local synchronization errors to dynamically reach a stabilizing gain without prior knowledge of the underlying network.}

%\rednotes{PS: Angela was proposing to call Prop 2 a Theorem so that we can show this more as a contribution wrt our 2026 ECC paper. What do you think?} \bluenotes{AR: I am not sure, but I have no problem with calling it a theorem.}
\begin{thm}\label{prop_adaptive}
Consider the closed-loop error system \eqref{eq:closed_loop} with the adaptive gain update law \eqref{eq:adaptiveupdatelaw}. Let Assumption \ref{assumption1} hold. If the auxiliary input $\phi(t, \bm \xi)$ is $u\delta$-PE, all signals in the closed-loop system ($\tilde{\bm x}(t), \tilde{\bm w}(t), c_1(t)$) are uniformly globally bounded. Furthermore, the system achieves simultaneous synchronization ($\lim_{t \to \infty} \tilde{\bm x}(t) = 0$) and directed topology identification ($\lim_{t \to \infty} \hat{\bm w}(t) =\bm  w$). 
\end{thm}

%\rednotes{PS: Here, either in the proof or as a remark, we can mention the following references that also use gain values depending on time or errors. So it is similar to here, also the Lyapunov functions, in some sense, can be similar.\\
%The work in \cite{li2014designing} proposes a controller whose gain is defined dynamically and grows monotonically as a function of the consensus errors. The work in \cite{LCSS-Maitryee} proposes a controller with a state-independent gain that grows slowly over time. In both studies, the coupling strength $c$ is redefined so that it no longer depends on global network information.\\
%You can change the phrasing etc btw, it is just a suggestion.}

\begin{proof}
Let $c^*$ be {an unknown,} sufficiently large positive constant such that $c^* > -\lambda_{\min}(L_{\mathrm{sym}})$. We define an augmented Lyapunov function that incorporates the estimation error of the adaptive gain:
\begin{equation*}
    V_{adapt} = \frac{1}{2}\tilde{\bm x}^\top \tilde{\bm x} + \frac{1}{2}\tilde{\bm w}^\top \tilde{\bm w} + \frac{1}{2\gamma_c}(c_1(t) - c^*)^2.
\end{equation*}
Taking the time derivative along the error trajectories, and incorporating the exact cancellation of the $\tilde{\bm w}$ cross-terms (as established in Proposition \ref{prop1} via the parameter update law $\dot{\hat{\bm w}} = -\Psi(\hat{\bm x})^\top \tilde{\bm x}$), we obtain:
\begin{equation*}
    \dot{V}_{adapt} = -\tilde{\bm x}^\top (c_1(t) I + L_{\mathrm{sym}})\tilde{\bm x} + \frac{1}{\gamma_c}(c_1(t) - c^*)\dot{c}_1(t).
\end{equation*}
Substituting the adaptation law \eqref{eq:adaptiveupdatelaw} into the derivative yields
\begin{equation*}
    \dot{V}_{adapt} = -c_1(t)\tilde{\bm x}^\top \tilde{\bm x} - \tilde{\bm x}^\top L_{\mathrm{sym}}\tilde{\bm x} + (c_1(t) - c^*)\tilde{\bm x}^\top \tilde{\bm x}.
\end{equation*}
The terms containing the time-varying $c_1(t)$ exactly cancel, simplifying the derivative to:
\begin{equation*}
    \dot{V}_{adapt} = -\tilde{\bm x}^\top (c^* I + L_{\mathrm{sym}})\tilde{\bm x}.
\end{equation*}
Because $c^*$ was chosen to dominate the instability of $L_{\mathrm{sym}}$, the matrix $(c^* I + L_{\mathrm{sym}})$ is strictly positive definite. Thus, there exists a scalar $\lambda > 0$ such that $\dot{V}_{adapt} \le -\lambda \lvert\tilde{\bm x}\rvert^2 \le 0$. 
%\rednotes{PS: Should we explicitly state smth like "Note that $c^*$ is introduced only for the proof to establish boundedness and is never computed, the adaptation law \eqref{eq:adaptiveupdatelaw} only depends on the synchronization error $\tilde x$."} \bluenotes{AR: I have added that \(c^*\) is unknown at the start of the proof.}

Since $\dot{V}_{adapt} \le 0$, the Lyapunov function $V_{adapt}$ is bounded, implying that $\tilde{\bm x}(t)$, $\tilde{\bm w}(t)$, and $c_1(t)$ are all uniformly globally bounded. Given that these signals and the state error derivative $\dot{\tilde{\bm x}}$ are bounded, $\tilde{\bm x}(t)$ is uniformly continuous. Integrating $\dot{V}_{adapt}$ over $[0, \infty)$ ensures $\tilde{\bm x} \in \mathcal{L}_2$. By Barbalat's Lemma, it follows that $\lim_{t \to \infty} \tilde{\bm x}(t) = 0$, guaranteeing synchronization. 

%The adaptive gain $c_1(t)$ monotonically increases until it provides sufficient damping to stabilize the divergent dynamics of the directed Laplacian. Finally, because $\hat{\bm x}(t)$ is $u\delta$-PE and $\tilde{\bm x}(t) \to 0$, standard persistency of excitation properties guarantee that the parameter estimation error converges to zero ($\lim_{t \to \infty} \tilde{\bm w}(t) = 0$).

The adaptive gain $c_1(t)$ monotonically increases and, being bounded, converges to a finite constant value that provides sufficient damping to stabilize the divergent dynamics of the directed Laplacian. To rigorously establish parameter convergence ($\lim_{t \to \infty} \tilde{\bm w}(t) = 0$), we analyze the asymptotic behavior of the state error dynamics \eqref{eq:error_dyn}. We have established that $\tilde{\bm x}(t) \to 0$ and that all signals in the closed-loop system are bounded. Because the derivative $\ddot{\tilde{\bm x}}(t)$ is a function of bounded variables ($\dot{\tilde{\bm x}}, \dot{c}_1, \dot{\Psi}$, and $\dot{\tilde{\bm w}}$), $\ddot{\tilde{\bm x}}(t)$ is bounded, making $\dot{\tilde{\bm x}}(t)$ uniformly continuous. By Barbalat's Lemma, this implies $\lim_{t \to \infty} \dot{\tilde{\bm x}}(t) = 0$.

%in \eqref{eq:error_dyn}
Taking the limit of the state error dynamics \(\dot{\tilde{\bm x}} = -(c_1 I + L)\tilde{\bm x} - \Psi(\hat{\bm x}(t))\tilde{\bm w}\)   as $t \to \infty$, and substituting both $\lim_{t \to \infty}\dot{\tilde{\bm x}}(t) = 0$ and $\lim_{t \to \infty} (c_1(t) I + L)\tilde{\bm x}(t) = 0$, yields:
%Taking the limit of the state error dynamics \eqref{eq:error_dyn} as $t \to \infty$, and noting that $(c_1(t) I + L)\tilde{\bm x}(t) \to 0$, yields:
%\bluenotes{PS: Here in Dimos's comment, I think it is written "Need to remind the reader why" for Eq. (14).} \rednotes{AR: In Eq. (10), you just plug in the values of \((c_1(t) I + L)\tilde{\bm x}(t)\) and \(\dot{\tilde{\bm x}}(t)\) which are zeros.. which then implies the below... not sure how else to write this.}
\begin{equation}
    \lim_{t \to \infty} \Psi(\hat{\bm x}(t))\tilde{\bm w}(t) = 0.
    \label{eq:psi_w_lim}
\end{equation}
Furthermore, taking the limit of the parameter update law \eqref{eq:update_law} gives:
\begin{equation}
    \lim_{t \to \infty} \dot{\tilde{\bm w}}(t) = \lim_{t \to \infty} \Psi(\hat{\bm x}(t))^\top \tilde{\bm x}(t) = 0,
    \label{eq:w_dot_lim}
\end{equation}
which implies $\tilde{\bm w}(\tau) = \tilde{\bm w}(t) + \varepsilon$ over any finite window $\tau \in [t, t+T]$ as $t \to \infty$ for a small positive \(\varepsilon\to 0^+\). Integrating $\lvert\Psi(\hat{\bm x}(\tau))\tilde{\bm w}(\tau)\rvert^2$ from $t$ to $t+T$ and using \eqref{eq:psi_w_lim} yields:
\begin{equation*}
    \lim_{t \to \infty} \tilde{\bm w}(t)^\top \left( \int_t^{t+T} \Psi(\hat{\bm x}(\tau))^\top \Psi(\hat{\bm x}(\tau)) d\tau \right) \tilde{\bm w}(t) = 0.
\end{equation*}
Since $\Psi(\hat{\bm x}(t))$ is $u\delta$-PE by Lemma~\ref{filprop}, the integral above is strictly positive definite ($\ge \mu I$) whenever $\lvert\tilde{\bm w}(t)\rvert \ge \delta > 0$, which forces $\lim_{t \to \infty} \tilde{\bm w}(t) = 0$. Thus, the directed topology is successfully identified.
\end{proof}

In practical applications, the PE signal $\phi(t, \bm \xi)$ cannot depend on the unmeasurable parameter error $\tilde{\bm w}$, as this requires prior knowledge of the true directed edge weights. To establish stability when the excitation signal relies solely on the measurable state synchronization error $\tilde{\bm x}$, we introduce a parameterized excitation signal $\phi_{\bm \theta}(t, \tilde{\bm x},{\bm \theta})$. Consequently, the auxiliary state $\hat{\bm x}(t)$ is now governed by the dynamics \eqref{eq:auxiliary_sys}, but with the excitation term $\phi(t, \tilde{\bm x})$ replaced by $\phi_{\bm \theta}(t, \tilde{\bm x}, {\bm \theta})$.
%\rednotes{PS: $\phi_{\bm \theta}(t, \tilde{\bm x}, {\bm \theta})?$ to show that it also depends on ${\bm \theta}$. Maybe it would also be good to refer to Eq. (8), saying now $\hat x$ is given by the update law (8) but with $\phi_{\bm \theta}$}. 
Here, ${\bm \theta} \in \mathbb{R}^{n_{\bm \theta}}$ represents a vector of constant tuning parameters (e.g., signal amplitudes and frequencies) that can be adjusted to arbitrarily shrink the ultimate tracking error. The dimension $n_{\bm \theta}$ is a control design choice determined by the network size $N$ which must be sufficiently large to accommodate the distinct frequency components required to satisfy the $u\delta$-PE condition for all unknown edge weights.

%In practical implementations, the persistent excitation (PE) signal $\phi_{\bm \theta}(t, \xi)$ cannot depend on the unmeasurable parameter error $\tilde{\bm w}$, as this requires a priori knowledge of the true directed edge weights. We prove that stability is preserved even when the excitation signal is restricted to depend solely on the measurable state synchronization error $\tilde{\bm x}$.

\begin{ass}\label{ass:regressor_bound}
Let $\rho: \mathbb{R}_{\ge0} \to \mathbb{R}_{\ge0}$ be a continuous non-decreasing function. The parameterized directed regressor matrix $\Psi_{\bm \theta}(t, \tilde{\bm x})$ satisfies the following bound for all $(t, \tilde{\bm x}) \in \mathbb{R}_{\ge 0} \times \mathbb{R}^N$:
\begin{equation*}
    \max \left\{ \lvert \Psi_{\bm \theta}(t,\tilde{\bm x})\rvert, \left\lvert \frac{\partial \Psi_{\bm \theta}(t,\tilde{\bm x})}{\partial t} \right\rvert, \left\lvert \frac{\partial \Psi_{\bm \theta}(t,\tilde{\bm x})}{\partial \tilde{\bm x}} \right\rvert \right\} \le \rho(\lvert\tilde{\bm x}\rvert).
\end{equation*}
\end{ass}
\begin{thm}
Consider the closed-loop error system \eqref{eq:closed_loop} with the adaptive gain $c_1(t)$ defined in \eqref{eq:adaptiveupdatelaw}, wherein the standard regressor $\Psi(\hat{\bm x})$ is replaced by the state-parameterized regressor $\Psi_{\bm \theta}(t, \tilde{\bm x})$. Let Assumption \ref{ass:regressor_bound} hold. If the regressor $\Psi_{\bm \theta}(t, \tilde{\bm x})$ is $u\delta$-PE with respect to $\tilde{\bm x}$, then the origin $(\tilde{\bm x}, \tilde{\bm w}) = (\bm 0,\bm 0)$ of the closed-loop system is uniformly semi-globally practically asymptotically stable.
\end{thm}
\begin{comment}
    \begin{thm}
Consider the closed-loop error system \eqref{eq:closed_loop} with $c_1$ as defined in \eqref{eq:adaptiveupdatelaw}, where the regressor $\Psi(\hat{\bm x})$ is replaced by the parameterized regressor $\Psi_{\bm \theta}(t, \tilde{\bm x})$. Let Assumption \ref{ass:regressor_bound} hold. If $\Psi_{\bm \theta}(t, \tilde{\bm x})$ is $u\delta$-PE with respect to $\tilde{\bm x}$, then the origin of the closed-loop error system $(\tilde{\bm x}, \tilde{\bm w}) = (0,0)$ is uniformly semiglobally practically asymptotically stable.
%Consider the control law and adaptive update laws defined in Section I. If the regressor matrix $\Psi_{\bm \theta}(t, \tilde{\bm x})$ is uniformly PE ($u\delta$-PE) with respect to $\tilde{\bm x}$, then the origin of the closed-loop error system $(\tilde{\bm x}, \tilde{\bm w}) = (0,0)$ is uniformly semiglobally practically asymptotically stable.
\end{thm}

\end{comment}
\begin{proof}
To establish stability without relying on $\tilde{\bm w}$ in the excitation signal, we construct a composite strict Lyapunov function:
\begin{equation}
    V(t,\bm  \xi, c_1(t)) = V_1(t, \bm \xi, c_1(t)) + \epsilon \big( V_2(t, \bm \xi) + V_3(t, \bm \xi) \big),
\end{equation}
where $\bm \xi = [\tilde{\bm x}^\top, \tilde{\bm w}^\top]^\top$, $\epsilon > 0$ is a sufficiently small positive scaling parameter, and the constituent functions are defined as:
\begin{subequations}
\begin{align}
    V_1 &= \frac{1}{2}\tilde{\bm x}^\top \tilde{\bm x} + \frac{1}{2}\tilde{\bm w}^\top \tilde{\bm w} + \frac{1}{2\gamma_c}(c_1(t) - c^*)^2, \\
    V_2 &= \tilde{\bm x}^\top \Psi_{\bm \theta}(t, \tilde{\bm x}) \tilde{\bm w}, \\
    V_3 &= -\tilde{\bm w}^\top M(t) \tilde{\bm w},
\end{align}
\end{subequations}
where $c^* > -\lambda_{\min}(L_{\mathrm{sym}})$ is the positive constant defined in the proof of Theorem~\ref{prop_adaptive}, $c_1(t)$ is governed by the adaptation law \eqref{eq:adaptiveupdatelaw}, and $M(t)$ is the matrix integral filter defined by:
\begin{equation*}
    M(t) = \int_t^\infty e^{(t-\tau)} \Psi_{\bm \theta}^\top(\tau, \tilde{\bm x})\Psi_{\bm \theta}(\tau, \tilde{\bm x}) d\tau.
\end{equation*}

Let $a_2 := c^* + \lambda_{\min}(L_{\mathrm{sym}}) > 0$. Following the analysis of Theorem~\ref{prop_adaptive} under the adaptation law \eqref{eq:adaptiveupdatelaw}, the time derivative of $V_1$ satisfies:
\begin{equation}
    \dot{V}_1 \le -a_2 \lvert\tilde{\bm x}\rvert^2,
    \label{eq:v1_dot}
\end{equation}
which ensures that $c_1(t)$ remains uniformly bounded. Consequently, there exists a constant $a_1 > 0$ such that $\lvert c_1(t) I + L \rvert = \lvert c_1(t) I + L^\top \rvert \le a_1$.

For a given compact domain $\mathbb{H}(\delta, \Delta) := \{ \bm \xi \in \mathbb{R}^{N+N(N-1)} : \delta \le \lvert\bm \xi\rvert \le \Delta \}$, note that $\lvert\tilde{\bm x}\rvert \le \Delta$ and $\lvert\tilde{\bm w}\rvert \le \Delta$. From Assumption~\ref{ass:regressor_bound}, since $\rho(\cdot)$ is non-decreasing, defining $b_\rho := \rho(\Delta) \ge 0$ yields $\lvert\Psi_{\bm \theta}(t, \tilde{\bm x})\rvert \le \rho(\lvert\tilde{\bm x}\rvert) \le b_\rho$. Furthermore, using the state error dynamics $\dot{\tilde{\bm x}} = -(c_1(t)I + L)\tilde{\bm x} - \Psi_{\bm \theta}(t, \tilde{\bm x})\tilde{\bm w}$, the state error derivative on $\mathbb{H}(\delta, \Delta)$ is bounded by:
\begin{equation*}
   \lvert\dot{\tilde{\bm x}}\rvert \le \lvert c_1(t)I + L \rvert \lvert\tilde{\bm x}\rvert + \lvert\Psi_{\bm \theta}\rvert \lvert\tilde{\bm w}\rvert \le (a_1 + b_\rho)\Delta.
\end{equation*}
Since $\dot{\Psi}_{\bm \theta}(t,\tilde{\bm x}(t)) = \frac{\partial \Psi_{\bm \theta}}{\partial t} + \frac{\partial \Psi_{\bm \theta}}{\partial \tilde{\bm x}}\dot{\tilde{\bm x}}$, it follows from Assumption~\ref{ass:regressor_bound} that $\lvert\dot{\Psi}_{\bm \theta}\rvert \le \rho(\lvert\tilde{\bm x}\rvert)\big(1+\lvert\dot{\tilde{\bm x}}\rvert\big) \le c_\rho$, where:
\begin{equation*}
   c_\rho := b_\rho\big(1+(a_1+b_\rho)\Delta\big).
\end{equation*}

\begin{comment}

For a given domain $\mathbb{H}(\delta, \Delta) := \{ \bm \xi \in \mathbb{R}^{N+N(N-1)} : \delta \le \lvert\bm \xi\rvert \le \Delta \}$, let $b_\rho := \rho(\Delta) \ge 0$. We define the strictly positive bounds:
\begin{equation*}
    \lvert\Psi_{\bm \theta}\rvert \le b_\rho, \quad \lvert c_1(t) I + L^\top\rvert \le a_1.
\end{equation*}
Since $\dot{\Psi}_{\bm \theta}(t,\tilde{\bm x}(t)) = \frac{\partial \Psi_{\bm \theta}}{\partial t} + \frac{\partial \Psi_{\bm \theta}}{\partial \tilde{\bm x}}\dot{\tilde{\bm x}}$, it follows from Assumption \ref{ass:regressor_bound} that $\lvert\dot{\Psi}_{\bm \theta}\rvert \le \rho(\lvert\tilde{\bm x}\rvert) + \rho(\lvert\tilde{\bm x}\rvert)\lvert\dot{\tilde{\bm x}}\rvert = \rho(\lvert\tilde{\bm x}\rvert)(1+\lvert\dot{\tilde{\bm x}}\rvert)$. On the domain $\mathbb{H}$, the state derivative is bounded by $\lvert\dot{\tilde{\bm x}}\rvert \le a_1\Delta + b_\rho\Delta = (a_1+b_\rho)\Delta$. Consequently, we obtain the bound $\lvert\dot{\Psi}_{\bm \theta}\rvert \le c_\rho$, where:
\begin{equation*}
    c_\rho := \rho(\Delta)\big(1+(a_1+b_\rho)\Delta\big).
\end{equation*}

Let $a_2 := c^* + \lambda_{\min}(L_{\mathrm{sym}}) > 0$. Following the analysis of Theorem \ref{prop_adaptive}, the time derivative of $V_1$ satisfies:
\begin{equation}
    \dot{V}_1 \le -a_2 \lvert\tilde{\bm x}\rvert^2.
    \label{eq:v1_dot}
\end{equation}
\end{comment}

Taking the time derivative of $V_2$ along the system trajectories yields:
\begin{equation*}
    \dot{V}_2 = \dot{\tilde{\bm x}}^\top \Psi_{\bm \theta} \tilde{\bm w} + \tilde{\bm x}^\top \dot{\Psi}_{\bm \theta} \tilde{\bm w} + \tilde{\bm x}^\top \Psi_{\bm \theta} \dot{\tilde{\bm w}}.
\end{equation*}
Substituting the modified error dynamics and the update law \eqref{eq:closed_loop} gives:
\begin{align*}
    \dot{V}_2 &= -\big((c_1(t) I + L)\tilde{\bm x} + \Psi_{\bm \theta} \tilde{\bm w}\big)^\top \Psi_{\bm \theta} \tilde{\bm w} \nonumber \\
    &\quad + \tilde{\bm x}^\top \dot{\Psi}_{\bm \theta} \tilde{\bm w} + \tilde{\bm x}^\top \Psi_{\bm \theta} \Psi_{\bm \theta}^\top \tilde{\bm x} \nonumber \\
    &= -\tilde{\bm x}^\top (c_1(t) I + L^\top)\Psi_{\bm \theta} \tilde{\bm w} -\lvert\Psi_{\bm \theta} \tilde{\bm w}\rvert^2 \nonumber \\
    &\quad + \tilde{\bm x}^\top \dot{\Psi}_{\bm \theta} \tilde{\bm w} + \lvert\Psi_{\bm \theta}^\top \tilde{\bm x}\rvert^2.
\end{align*}
Applying the Cauchy--Schwarz inequality with the bounds $\lvert c_1(t)I + L^\top \rvert \le a_1$, $\lvert\Psi_{\bm \theta}\rvert \le b_\rho$, and $\lvert\dot{\Psi}_{\bm \theta}\rvert \le c_\rho$, and defining the constant $K_2 := a_1 b_\rho + c_\rho$, we obtain:
%Applying the Cauchy-Schwarz inequality and defining the constant $K_2 := a_1 b_\rho + c_\rho$, we obtain the bound:
\begin{equation}
    \dot{V}_2 \le -\lvert\Psi_{\bm \theta} \tilde{\bm w}\rvert^2 + b_\rho^2 \lvert\tilde{\bm x}\rvert^2 + K_2 \lvert\tilde{\bm x}\rvert \lvert\tilde{\bm w}\rvert.
    \label{eq:v2_dot}
\end{equation}

Taking the time derivative of $V_3$ using the product rule yields:
\begin{equation}
    \dot{V}_3 = -2\dot{\tilde{\bm w}}^\top M(t) \tilde{\bm w} - \tilde{\bm w}^\top \dot{M}(t) \tilde{\bm w}.
    %\dot{V}_3 = -2\tilde{\bm w}^\top \dot{\tilde{\bm w}} M(t) - \tilde{\bm w}^\top \dot{M}(t) \tilde{\bm w}.
    \label{eq:V3dot}
\end{equation}
Applying the Leibniz integral rule, the time derivative of the integral matrix $M(t)$ evaluates to:
\begin{equation*}
    \dot{M}(t) = -\Psi_{\bm \theta}^\top(t,\tilde{\bm x})\Psi_{\bm \theta}(t,\tilde{\bm x}) + M(t).
\end{equation*}
Substituting $\dot{M}(t)$ and the update law $\dot{\tilde{\bm w}} = \Psi_{\bm \theta}^\top \tilde{\bm x}$ from \eqref{eq:closed_loop} into \eqref{eq:V3dot} provides:
%Substituting $\dot{M}(t)$ and the update law $\dot{\tilde{\bm w}}$ \eqref{eq:closed_loop} back into \eqref{eq:V3dot} provides:
\begin{align*}
    \dot{V}_3 &= -2\tilde{\bm x}^\top \Psi_{\bm \theta} M(t) \tilde{\bm w} - \tilde{\bm w}^\top \big(-\Psi_{\bm \theta}^\top \Psi_{\bm \theta} + M(t)\big) \tilde{\bm w} \nonumber \\
    &= -2\tilde{\bm x}^\top \Psi_{\bm \theta} M(t) \tilde{\bm w} + \lvert\Psi_{\bm \theta} \tilde{\bm w}\rvert^2 - \tilde{\bm w}^\top M(t) \tilde{\bm w}.
\end{align*}
Because $\lvert M(t) \rvert \le b_\rho^2 \int_t^\infty e^{(t-\tau)} d\tau = b_\rho^2$ on $\mathbb{H}(\delta, \Delta)$, we can bound the cross-term by defining $K_3 := 2 b_\rho^3$, yielding:
%Because $M(t)$ is bounded by $b_\rho^2 I$ on the defined domain, we can bound the cross-term by defining $K_3 := 2 b_\rho^3$, yielding:
\begin{equation}
    \dot{V}_3 \le \lvert\Psi_{\bm \theta} \tilde{\bm w}\rvert^2 - \tilde{\bm w}^\top M(t) \tilde{\bm w} + K_3 \lvert\tilde{\bm x}\rvert \lvert\tilde{\bm w}\rvert.
    \label{eq:v3_dot}
\end{equation}
Summing \eqref{eq:v1_dot}, \eqref{eq:v2_dot}, and \eqref{eq:v3_dot}, the indefinite terms $-\lvert\Psi_{\bm \theta} \tilde{\bm w}\rvert^2$ and $+\lvert\Psi_{\bm \theta} \tilde{\bm w}\rvert^2$ cancel. Furthermore, since $e^{(t-\tau)} \ge e^{-T}$ for all $\tau \in [t, t+T]$ and the regressor $\Psi_{\bm \theta}(t, \tilde{\bm x})$ is $u\delta$-PE (Definition~\ref{delta-PE}), the matrix $M(t)$ is strictly bounded from below on $\mathbb{H}(\delta, \Delta)$ over any window $T > 0$ by:
%Summing \eqref{eq:v1_dot}, \eqref{eq:v2_dot}, and \eqref{eq:v3_dot}, the indefinite boundary terms $-\lvert\Psi_{\bm \theta} \tilde{\bm w}\rvert^2$ and $+\lvert\Psi_{\bm \theta} \tilde{\bm w}\rvert^2$  cancel. Because the directed regressor $\Psi_{\bm \theta}(t, \tilde{\bm x})$ is assumed to be $u\delta$-PE, the integral $M(t)$ is strictly bounded from below by a positive definite matrix over any window $T$:
\begin{equation*}
    M(t) \ge \int_t^{t+T} e^{(t-\tau)} \Psi_{\bm \theta}^\top \Psi_{\bm \theta} d\tau \ge \mu I e^{-T}.
\end{equation*}

%\rednotes{Since \(\tau\in[t,t+T]\), we have \(e^{t-\tau}\geq e^{-T}\), So, \(M(t)\geq e^{-T}\int_t^{t+T}  \Psi_{\bm \theta}^\top \Psi_{\bm \theta} d\tau\) after which applying definition 2, gives us \(e^{-T}\int_t^{t+T}  \Psi_{\bm \theta}^\top \Psi_{\bm \theta} d\tau\ge \mu I e^{-T}\).}

%\rednotes{AR: Need to merge the above properly.}

Letting $b' := \mu e^{-T} > 0$, we establish a strictly negative definite bound for the unmeasurable parameter error:
\begin{equation*}
    -\tilde{\bm w}^\top M(t) \tilde{\bm w} \le -b' \lvert\tilde{\bm w}\rvert^2.
\end{equation*}

Letting $K := K_2 + K_3$, the derivative of the composite Lyapunov function becomes:
\begin{equation*}
    \dot{V} \le -a_2 \lvert\tilde{\bm x}\rvert^2 + \epsilon b_\rho^2 \lvert\tilde{\bm x}\rvert^2 - \epsilon b' \lvert\tilde{\bm w}\rvert^2 + \epsilon K \lvert\tilde{\bm x}\rvert \lvert\tilde{\bm w}\rvert.
\end{equation*}
Applying Young's Inequality ($ab \le \frac{\eta}{2}a^2 + \frac{1}{2\eta}b^2$ for any $\eta > 0$) to the indefinite cross-term gives:
\begin{equation*}
    \epsilon K \lvert\tilde{\bm x}\rvert \lvert\tilde{\bm w}\rvert \le \epsilon \left( \frac{\eta}{2} K \lvert\tilde{\bm x}\rvert^2 + \frac{1}{2\eta} K \lvert\tilde{\bm w}\rvert^2 \right).
\end{equation*}
Grouping the corresponding error state variables yields:
\begin{equation*}
    \dot{V} \le -\left(a_2 - \epsilon b_\rho^2 - \epsilon \frac{\eta}{2} K \right)\lvert\tilde{\bm x}\rvert^2 - \epsilon\left(b' - \frac{1}{2\eta} K \right)\lvert\tilde{\bm w}\rvert^2.
\end{equation*}
By choosing $\eta$ sufficiently large such that $b' > \frac{1}{2\eta} K$, and choosing $\epsilon > 0$ sufficiently small such that $a_2 > \epsilon(b_\rho^2 + \frac{\eta}{2} K)$, both grouped coefficients become strictly positive scalars $\alpha > 0$ and $\beta > 0$. Thus:
\begin{equation*}
    \dot{V} \le -\alpha \lvert\tilde{\bm x}\rvert^2 - \beta \lvert\tilde{\bm w}\rvert^2 \le 0.
\end{equation*}
By Chaillet Theorem \cite{chaillet2008uniform}, the proof is complete and the origin of the closed-loop system is uniformly semiglobally practically asymptotically stable.
\end{proof}

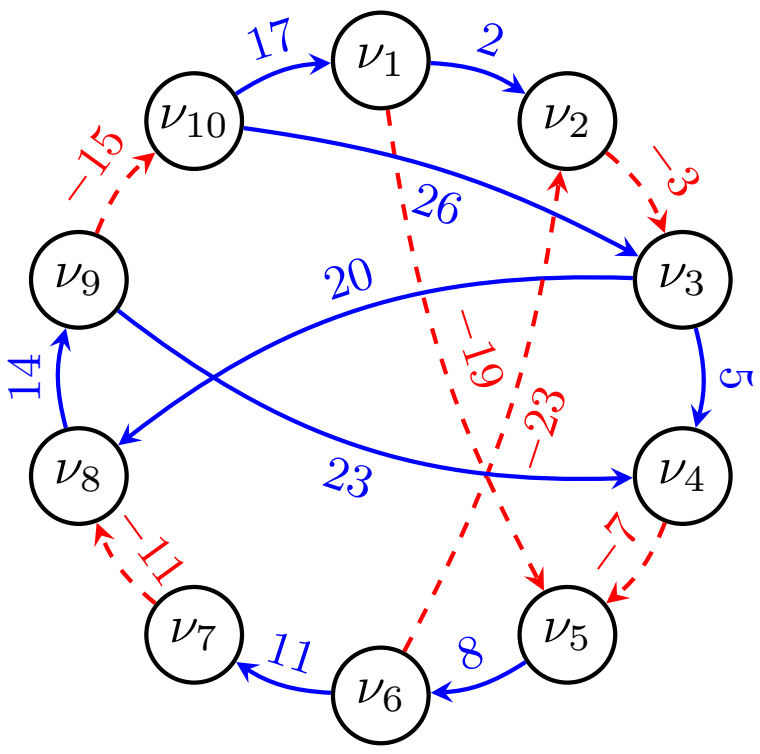
\begin{figure}[!t]
\centering
\begin{tikzpicture}[thick,scale=0.6,
  mynode/.style={circle,draw,minimum size=18pt,inner sep=0pt,thick},
  comp/.style={->, >=stealth, blue, thick},
  antag/.style={->, >=stealth, red, thick, dashed},
  % New style to keep text readable, sloped alongside the line, and hovering off the edge
  elabel/.style={midway, font=\small, sloped, inner sep=3pt}
  ]

% ==== 10 Nodes arranged in a circle ====
\foreach \i [count=\n from 1] in {90,54,18,-18,-54,-90,-126,-162,-198,-234}{
    \node[mynode] (\n) at ({3.5*cos(\i)},{3.5*sin(\i)}) {$\nu_{\n}$};
}

% ==== Sparse Directed Edges with Spaced Weights ====
% comp = positive (blue solid), antag = negative (red dashed)

% Ring edges (placed "above" the sloped path, which pushes them to the outside of the ring)
\draw[comp] (1) to[bend left=15] node[elabel, above] {$2$} (2);
\draw[antag] (2) to[bend left=15] node[elabel, above] {$-3$} (3);
\draw[comp] (3) to[bend left=15] node[elabel, above] {$5$} (4);
\draw[antag] (4) to[bend left=15] node[elabel, above] {$-7$} (5);
\draw[comp] (5) to[bend left=15] node[elabel, above] {$8$} (6);
\draw[comp] (6) to[bend left=15] node[elabel, above] {$11$} (7);
\draw[antag] (7) to[bend left=15] node[elabel, above] {$-11$} (8);
\draw[comp] (8) to[bend left=15] node[elabel, above] {$14$} (9);
\draw[antag] (9) to[bend left=15] node[elabel, above] {$-15$} (10);
\draw[comp] (10) to[bend left=15] node[elabel, above] {$17$} (1);

% Internal cross edges (mixing "above" and "below" to utilize open space efficiently)
\draw[antag] (1) to[bend right=10] node[elabel, above] {$-19$} (5);
\draw[comp] (3) to[bend right=20] node[elabel, above] {$20$} (8);
\draw[antag] (6) to[bend right=10] node[elabel, below] {$-23$} (2);
\draw[comp] (9) to[bend right=20] node[elabel, below] {$23$} (4);
\draw[comp] (10) to[bend left=10] node[elabel, below] {$26$} (3);

\end{tikzpicture}
\caption{The sparse directed signed graph of the $10$ agents. }
\label{fig_top}
\end{figure}

\begin{figure}[!t]
    \centering
    % (a) State Synchronization Error
    \subfloat[State synchronization errors $\tilde{ x}_i(t)$]{
        \includegraphics[width=1\linewidth]{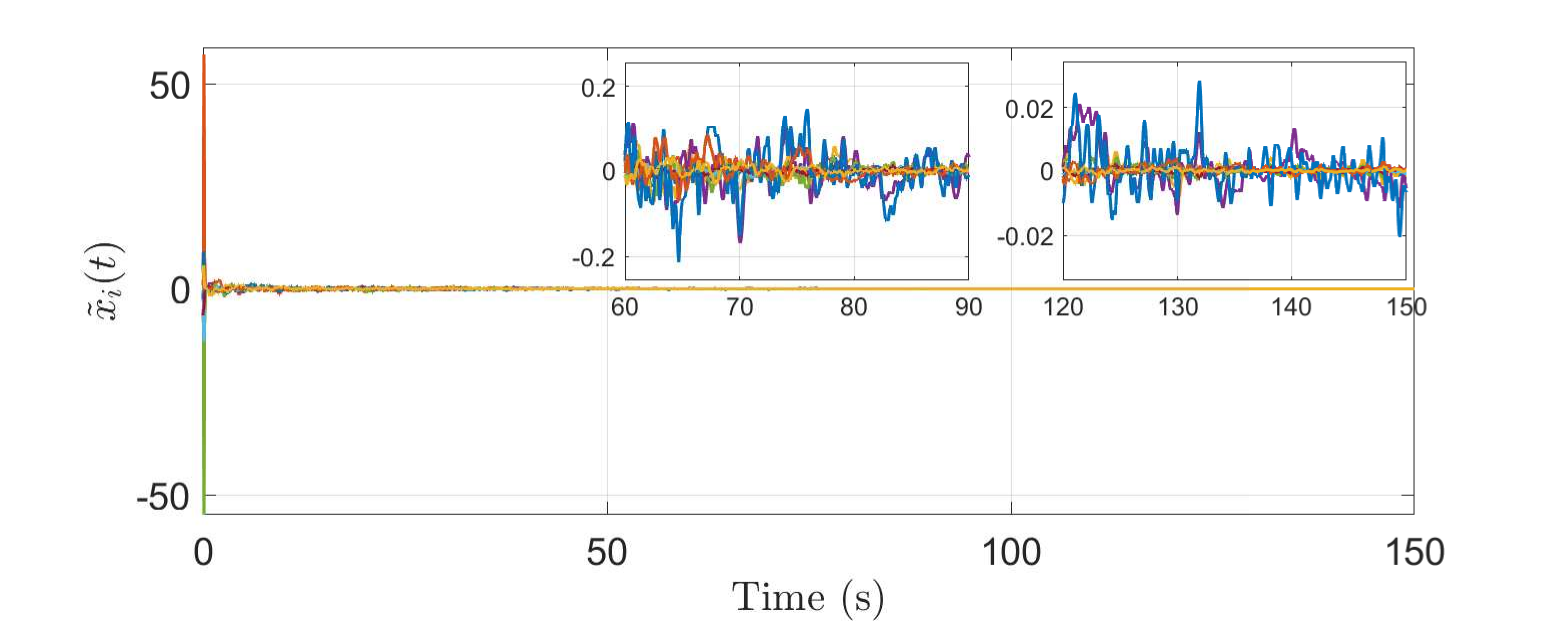}
        \label{fig:statesync}
    }
    
   % \vspace{0.2cm} % Adds slight vertical spacing between plots
    
    % (b) Parameter Estimation Error
    \subfloat[Parameter estimation errors $\tilde{ w}_k(t)$]{
        \includegraphics[width=1\linewidth]{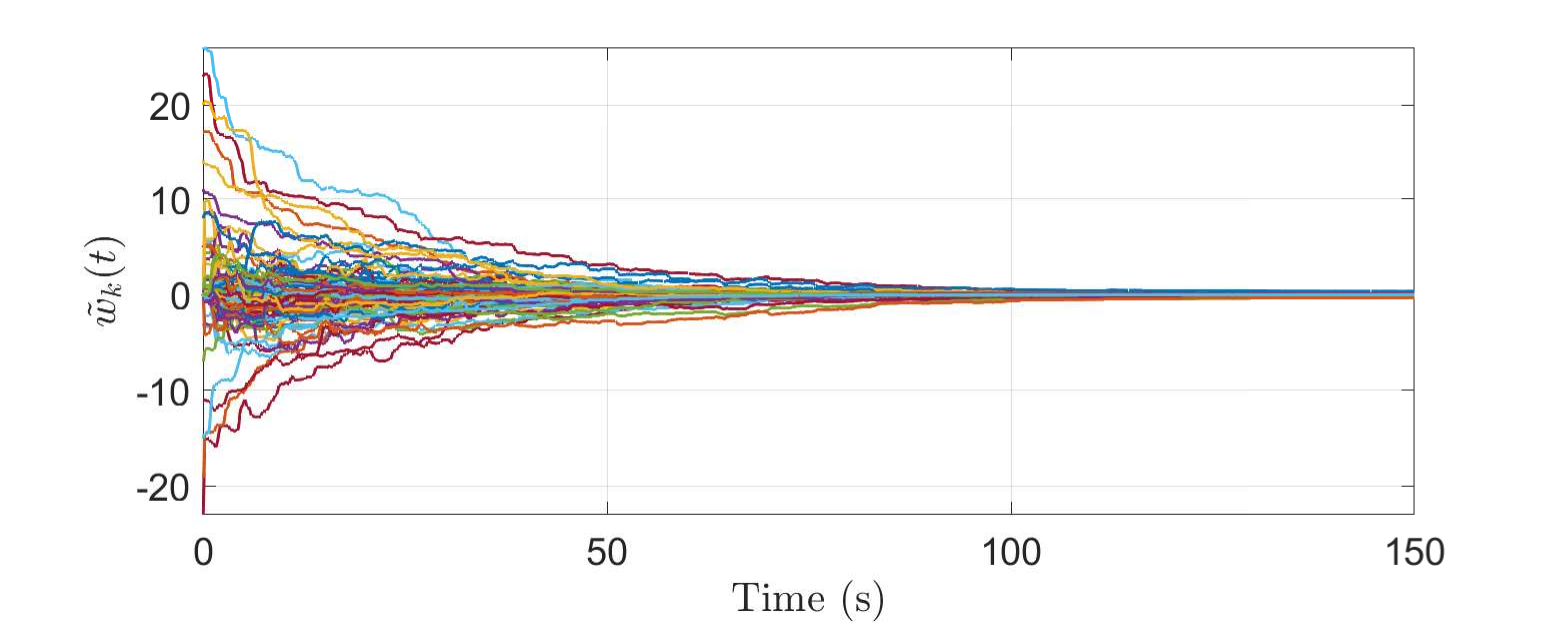}
        \label{fig:paramerror}
    }
    
   % \vspace{0.2cm}
    
    % (c) Estimated Trajectories
    \subfloat[Trajectories of identified directed weights $\hat{w}_k(t)$]{
        \includegraphics[width=1\linewidth]{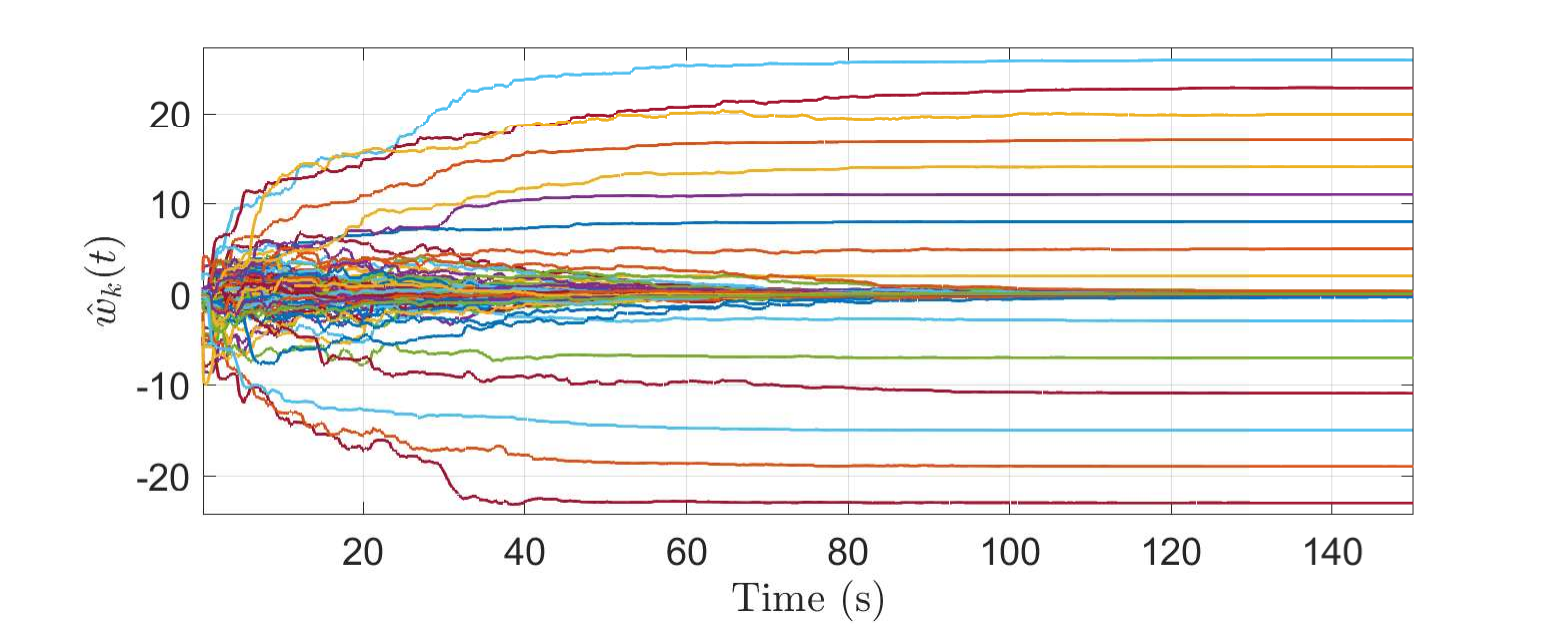}
        \label{fig:estimatedtraj}
    }
    
    \caption{Evolution of the trajectories.}
    \label{fig:sim_results}
\end{figure}

\section{Numerical Example}
%\rednotes{PS: I think here you can also put the value you chose for $\gamma_c$ and the persistently excited signal. Yes yes no worries Ahsan, I just put stuff this week because I will be a bit busy next week so I will probably have less time, so no pressure.}

%\bluenotes{A bit busy in italy.. I will complete this next week. I remember that the persistently excited signal had 30 frequencies.. maybe I will decrease them}

We consider a network of $N=10$ agents modeled by \eqref{eq:systemequation} interconnected over an unknown, sparse, directed signed network, represented in Figure \ref{fig_top}. The objective is to simultaneously synchronize the network and identify the $N(N-1)=90$ unknown directed signed weights $w_k$. Assuming no prior global knowledge of the network topology or the spectrum of $L_{\mathrm{sym}}$, we deploy the adaptive control protocol \eqref{eq:controlinput} with the parameter update law \eqref{eq:update_law}, the auxiliary filter \eqref{eq:auxiliary_sys}, and the adaptive gain update law \eqref{eq:adaptiveupdatelaw}. % $\dot{c}_1(t) = \gamma_c \|\tilde{x}(t)\|^2$.

To ensure the excitation relies exclusively on measurable local state errors, the input for each agent $i \in \{1, \dots, 10\}$ is chosen as $\phi_{\bm \theta, i}(t, \tilde{x}_i) = \tanh(\kappa \tilde{x}_i) p_{e,i}(t)$ with $\kappa = 1000$, where the excitation signal $p_{e,i}(t)$ is constructed as a sum of sinusoids inspired by \cite{sekercioglu2026topology}:

%The objective is to identify the $N(N-1)=90$ unknown directed signed weights $w_{k}$. 
%To ensure the excitation relies exclusively on measurable local state errors, the input is chosen as $\phi_{\bm \theta}(t, \tilde{\bm x}) = \tanh(\kappa \tilde{\bm x}) \circ \bm p_e(t)$ with $\kappa = 1000$, where $\circ$ denotes the Hadamard product. For each agent $i \in \{1, \dots, 10\}$, the excitation signal $p_{e,i}(t)$ is constructed using $5$ shared frequencies:
\begin{equation*}
    p_{e,i}(t) = \sum_{m=1}^{5} a_m \sin\left(\omega_m t + \frac{2\pi(i-1)m}{N+1}\right),
\end{equation*}
with amplitudes $\bm a = [3, 4, 5, 6, 7]^\top$ and frequencies $\bm \omega = [2, 3, 5, 7, 11]^\top\text{ rad/s}$. The learning rate is set to $\gamma_c = 0.1$ to maintain sufficient excitation during the parameter estimation transient, with initial conditions $c_1(0) = 0.1$, $\hat{\bm x}(0) = \bm 0$, and $\hat{\bm w}(0) = \bm 0$.

%To prevent the adaptive gain $c_1(t)$ from prematurely suppressing $\tilde{\bm x}(t)$ and extinguishing $\phi_{\bm \theta}(t, \tilde{\bm x})$ before all $90$ parameters converge, we select $\gamma_c = 0.1$ with $c_1(0) = 0.1$, $\hat{\bm x}(0) = 0$, and $\hat{\bm w}(0) = 0$.

The simulation results are presented in Figure~\ref{fig:sim_results}. As shown in Figure~\ref{fig:statesync}, the state synchronization errors $\tilde{x}_i(t)$ converge to a small neighborhood of the origin. Simultaneously, the weight estimation errors $\tilde{w}_k(t)$ converge to a small neighborhood of zero (Figure~\ref{fig:paramerror}), indicating that the directed signed weights are successfully identified (Figure~\ref{fig:estimatedtraj}). Although the estimation errors do not converge to exact zero, the residual neighborhood can be made arbitrarily small by tuning $\kappa$. Finally, the proposed strategy achieves simultaneous topology identification and synchronization without requiring global knowledge of the network.

\section{CONCLUSIONS}
We presented an adaptive control framework for simultaneous topology identification and synchronization of directed signed networks modeled by repelling Laplacians. Using a node-centric regressor and a $u\delta$-PE auxiliary system, we established uniform global asymptotic stability of the closed-loop error system under known spectral bounds, and subsequently removed the need for global network knowledge via a time-varying adaptive gain. Finally, we proved uniform semiglobal practical asymptotic stability when the excitation depends solely on measurable synchronization errors, and validated the approach through numerical simulations. 

%\rednotes{PS: I added one paper with Nana as first author, removed on paper of Claudio, and one paper of Esteban-Nana-Dimos. What do you think we can do more for less self-citations?}
%\rednotes{AR: At the moment, we have 7 out of 18 self citations... maybe we increase the total number of references.. and have some more general references.. maybe in the proofs or before in the introduction.. I did not know where to find the reference for the fundamental property that dimos asked in theorem 1? So, I proved it instead. maybe if you know where to refer it from, we can change it back and add the references.?} \bluenotes{PS: Let's also use this ref while talking about matrix or algebraic stuff: \cite{Horn}. Like in Prop 1 and later on.}
\bibliographystyle{IEEEtran}
\bibliography{references_pelin,reference_nana} %,references_angela

@book{horn2012matrix,
  title={Matrix analysis},
  author={Horn, Roger A and Johnson, Charles R},
  year={2012},
  publisher={Cambridge university press}
}

@article{fontan2022multiagent,
  title={Multiagent consensus over time-invariant and time-varying signed digraphs via eventual positivity},
  author={Fontan, Angela and Wang, Lingfei and Hong, Yiguang and Shi, Guodong and Altafini, Claudio},
  journal={IEEE Transactions on Automatic Control},
  volume={68},
  number={9},
  pages={5429--5444},
  year={2022},
  publisher={IEEE}
}

@INPROCEEDINGS{sekercioglu2026topology,
  author={Sekercioglu, Pelin and Wang, Nana and Fontan, Angela and Dimarogonas, Dimos V.},
  booktitle={European Control Conference}, 
  title={Topology Identification of Dynamical Signed Graphs}, 
  year={2026},
  pages={3321-3326}}

@article{Fontan2021Signed,
author = {Fontan, Angela and Altafini, Claudio},
doi = {10.1038/s41598-021-84147-3},
issn = {2045-2322},
journal = {Scientific Reports},
number = {5134},
pmid = {33664333},
publisher = {Nature Publishing Group UK},
title = {{A signed network perspective on the government formation process in parliamentary democracies}},
volume = {11},
year = {2021}
}

@preamble{"\def\nonumero{\def\numerodeitem{}}"}

@string{SIAM = "SIAM J. on Contr. and Opt."}

@preamble{ "\def\nonumero{\def\numerodeitem{}}" }

@book{KHALIL92,
	title        = {Nonlinear systems},
	author       = {H. Khalil},
	year         = 1992,
	publisher    = {{Macmillan Publishing Co., 1st ed.}},
	address      = {New York}
}

@ARTICLE{altafini2012_6329411,
	title        = {Consensus Problems on Networks With Antagonistic Interactions},
	author       = {C. Altafini},
	year         = 2013,
	journal      = {IEEE Trans. on Autom. Cont.},
	volume       = 58,
	number       = 4,
	pages        = {935--946},
	doi          = {10.1109/TAC.2012.2224251}
}

@Article{LCSS-Maitryee,
  author = 	 {M. Dutta and E. Panteley and A. Lor\'{\i}a and S. Sukumar},
  title = 	 {Strict {Lyapunov} functions for dynamic consensus in linear systems interconnected over directed graphs},
  journal	= {IEEE Control Systems Letters},
  year = 	 2022,
  volume = 	 6,
  number = 	 1,
  pages = 	 {2323--2328},
  doi = 	 {10.1109/LCSYS.2022.3150662},
  hal_id = {hal-03752255}
}

@article{restrepo2023simultaneousTCNS,
  title={Simultaneous topology identification and synchronization of directed dynamical networks},
  author={Restrepo, Esteban and Wang, Nana and Dimarogonas, Dimos V},
  journal={IEEE Trans. on Cont. of Network Sys.},
  volume={11},
  number={3},
  pages={1491--1501},
  year={2023}
}

@inproceedings{loria1999new,
  title={A new notion of persistency-of-excitation for UGAS of NLTV systems: Application to stabilisation of nonholonomic systems},
  author={Lor{\'\i}a, Antonio and Panteley, Elena and Teel, Andrew},
  booktitle={Eur. Control Conf.},
  pages={1363--1368},
  year={1999}
}

@article{panteley2001relaxed,
  title={Relaxed persistency of excitation for uniform asymptotic stability},
  author={Panteley, Elena and Loria, Antonio and Teel, Andrew},
  journal={IEEE Trans. on Autom. Cont.},
  volume={46},
  number={12},
  pages={1874--1886},
  year={2001}
}

@article{li2014designing,
  title={Designing fully distributed consensus protocols for linear multi-agent systems with directed graphs},
  author={Li, Zhongkui and Wen, Guanghui and Duan, Zhisheng and Ren, Wei},
  journal={IEEE Trans. on Autom. Cont.},
  volume={60},
  number={4},
  pages={1152--1157},
  year={2014}
}

@article{newman2003structure,
  title={The structure and function of complex networks},
  author={Newman, Mark EJ},
  journal={SIAM review},
  volume={45},
  number={2},
  pages={167--256},
  year={2003},
  publisher={SIAM}
}

@ARTICLE{11045067,
  author={Gogianu, Florin and Buşoniu, Lucian and Morărescu, Irinel-Constantin},
  journal={IEEE Control Systems Letters}, 
  title={Learning-Based Control of the Consensus Value in Unknown Graphs}, 
  year={2025},
  volume={9},
  pages={1093-1098}}

@article{chaillet2008uniform,
  title={Uniform semiglobal practical asymptotic stability for non-autonomous cascaded systems and applications},
  author={Chaillet, Antoine and Lor{\'\i}a, Antonio},
  journal={Automatica},
  volume={44},
  number={2},
  pages={337--347},
  year={2008},
  publisher={Elsevier}
}

@article{zhu_new_2021,
	title = {A {New} {Method} for {Topology} {Identification} of {Complex} {Dynamical} {Networks}},
	volume = {51},
	doi = {10.1109/TCYB.2019.2894838},
	number = {4},
	journal = {IEEE Trans. on Cybernetics},
	author = {Zhu, Shuaibing and Zhou, Jin and Chen, Guanrong and Lu, Jun-An},
	year = {2021},
	pages = {2224--2231}
}

@article{van_waarde_topology_2019,
	title = {Topology {Reconstruction} of {Dynamical} {Networks} via {Constrained} {Lyapunov} {Equations}},
	volume = {64},
	number = {10},
	urldate = {2025-01-24},
	journal = {IEEE Trans. on Autom. Cont.},
	author = {Van Waarde, Henk J. and Tesi, Pietro and Camlibel, M. Kanat},
	month = oct,
	year = {2019},
	pages = {4300--4306}
}

@article{weerts2018prediction,
  title={Prediction error identification of linear dynamic networks with rank-reduced noise},
  author={Weerts, Harm HM and Van den Hof, Paul MJ and Dankers, Arne G},
  journal={Automatica},
  volume={98},
  pages={256--268},
  year={2018},
  publisher={Elsevier}
}

@inproceedings{wang2023finite,
  title={Finite-time topology identification for complex dynamical networks},
  author={Wang, Nana and Dimarogonas, Dimos V},
  booktitle={62nd IEEE Conf. on Dec. and Cont.},
  pages={425--430},
  year={2023},
}

@article{IEEE-robotmanip,
  title={Robust Formation Control of Robot Manipulators with Inter-agent Constraints over Undirected Signed Networks},
  author={{\c{S}}ekercio{\u{g}}lu, P. and Jayawardhana, B. and Sarras, I. and Lor{\'\i}a, A. and Marzat, J.},
  journal={IEEE Trans. on Cont. of Network Sys.},
    year={2025},
  volume={12},
  number={1},
  pages={251-261}
}
\end{document}